\documentclass[sigconf]{acmart}
\usepackage{float}
\usepackage{multirow}
\usepackage{CJKutf8}
\usepackage{booktabs}
\usepackage{graphicx}
\usepackage{amsmath}
\usepackage{algorithmic}
\usepackage[ruled,linesnumbered]{algorithm2e}
\usepackage{amsfonts}
\usepackage[justification=centering]{caption}
\usepackage{makecell}
\usepackage{enumitem}
\usepackage{float}
\usepackage{subfig}
\usepackage{balance}
\AtBeginDocument{%
  }

\copyrightyear{2026}
\acmYear{2026}
\setcopyright{cc}
\setcctype{by}
\acmConference[KDD '26]{Proceedings of the 32nd ACM SIGKDD Conference on Knowledge Discovery and Data Mining V.1}{August 09--13, 2026}{Jeju Island, Republic of Korea}
\acmBooktitle{Proceedings of the 32nd ACM SIGKDD Conference on Knowledge Discovery and Data Mining V.1 (KDD '26), August 09--13, 2026, Jeju Island, Republic of Korea}
\acmPrice{}
\acmDOI{10.1145/3770854.3780177}
\acmISBN{979-8-4007-2258-5/2026/08}

\begin{document}

\title{FCN: Fusing Exponential and Linear Cross Network for Click-Through Rate Prediction}

\author{Honghao Li}
\email{salmon1802li@gmail.com}
\orcid{0009-0000-6818-7834}
\affiliation{%
  \institution{Anhui University}
  \city{Hefei}
  \state{Anhui Province}
  \country{China}
}

\author{Yiwen Zhang}
\authornote{Corresponding author}
\email{zhangyiwen@ahu.edu.cn}
\affiliation{%
  \institution{Anhui University}
  \city{Hefei}
  \state{Anhui Province}
  \country{China}
}

\author{Yi Zhang}
\email{zhangyi.ahu@gmail.com}
\affiliation{%
  \institution{Anhui University}
  \city{Hefei}
  \state{Anhui Province}
  \country{China}
}

\author{Hanwei Li}
\email{lihanwei@stu.ahu.edu.cn}
\affiliation{%
  \institution{Anhui University}
  \city{Hefei}
  \state{Anhui Province}
  \country{China}
}

\author{Lei Sang}
\email{sanglei@ahu.edu.cn}
\affiliation{%
  \institution{Anhui University}
  \city{Hefei}
  \state{Anhui Province}
  \country{China}
}

\author{Jieming Zhu}
\email{jiemingzhu@ieee.org}
\affiliation{%
  \institution{Huawei Noah’s Ark Lab}
  \city{Shenzhen}
  \state{Guangdong Province}
  \country{China}
}

\renewcommand{\shortauthors}{Li et al.}

\begin{abstract}
  As an important modeling paradigm in click-through rate (CTR) prediction, the Deep \& Cross Network and its derivative models have gained widespread recognition, primarily due to their success in trade-off computational cost and performance. However, this paradigm typically depends on deep neural network (DNN) to implicitly learn high-order feature interactions, without explicitly modeling extremely high-order interactions due to concerns about model complexity. To address this limitation, we propose a novel model for CTR prediction, called the Fusing Cross Network (FCN), which consists of two sub-networks: the Exponential Cross Network (ECN) and the Linear Cross Network (LCN). Specifically, ECN explicitly captures extremely high-order feature interactions whose order increases exponentially with network depth, while LCN captures low-order feature interactions with linearly increasing order. By integrating these two sub-networks, FCN is able to explicitly model a broad spectrum of feature interactions, thereby eliminating the need to rely on implicit modeling by DNN. Moreover, we introduce a low-cost aggregation method that reduces the number of parameters by 50\% and inference latency by 23\%. Meanwhile, we propose a simple yet effective loss function, Tri-BCE, which provides tailored supervision signals for each sub-network. We evaluate the effectiveness and efficiency of FCN on six public benchmark datasets and 16 baselines. Furthermore, we verify the effectiveness of the FCN on a real-world business dataset spanning seven days. The code, running logs, and detailed hyperparameter configurations are publicly available at \url{https://github.com/salmon1802/FCN}.
\end{abstract}




\maketitle

\begin{figure}[t]
  \centering
  \includegraphics[width=0.48\textwidth]{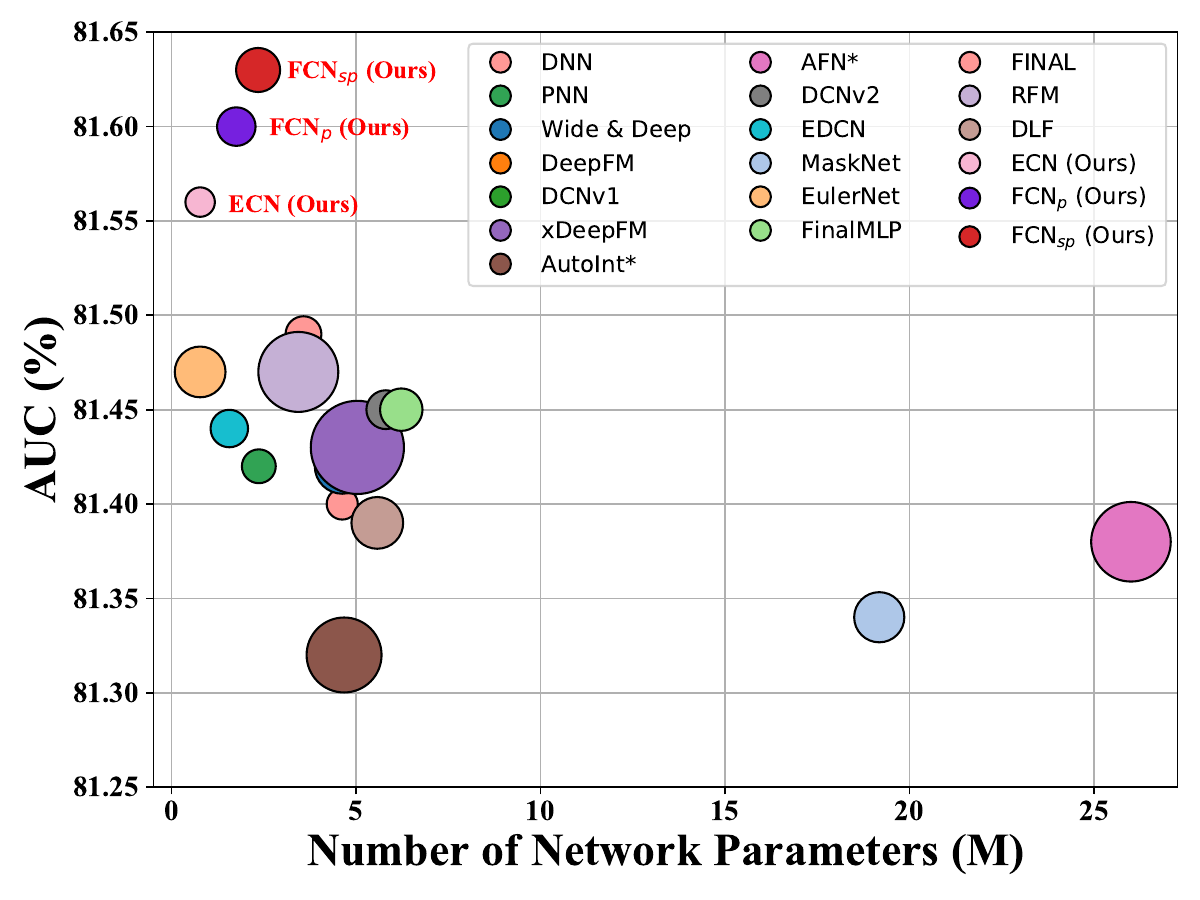}
  \captionsetup{justification=justified, singlelinecheck=false}
  \caption{ Comparison among ECN, FCN, and other models in terms of network parameter number, AUC, and running time on Criteo dataset. The graphic area represents the running time per epoch for each model (a larger area indicates a longer time, and vice versa).}
  \label{benchmark}
\end{figure}

\section{Introduction}
\label{Introduction}
Click-through rate (CTR) prediction is an essential part of industrial recommender systems and online advertising \cite{openbenchmark, Bars, MEGG, xianquan1}. It uses heterogeneous features as model inputs, such as user profiles, item attributes, and context \cite{CETN, FINAL}. These features undergo interaction modeling to predict the probability that a user will click on an item, thereby providing a better user experience and increasing the profitability of the recommender system \cite{dcnv2, EDCN, deepfm}.

As a representative feature interaction-based CTR modeling paradigm, Deep \& Cross Network (DCN)~\cite{dcn, dcnv2} achieves a favorable trade-off between computational cost and model performance, attracting considerable attention from CTR researchers~\cite{xdeepfm, autoint, AFN, EulerNet}. As its name suggests, DCN is a "deep and cross" network rather than a "deep cross" network, as it consists of both a deep neural network (DNN) and a cross network (CrossNet). In this architecture, the DNN is responsible for modeling implicit high-order feature interactions, while the CrossNet explicitly captures low-order feature interactions, typically only up to the third or fourth order~\cite{xdeepfm, openbenchmark}. Nevertheless, several studies~\cite{FINAL, GDCN, autofis} indicate that high-order feature interactions are beneficial for model performance, but modeling extremely high-order interactions remains challenging due to gradient \cite{gradient_explosion_vanishment}, rank \cite{rank_low}, and noise \cite{autofis} issues. Furthermore, with the establishment of open-source benchmarks~\cite{openbenchmark, Bars}, researchers observe that CTR models appear to encounter a performance bottleneck: when well-tuned hyperparameters are used, the performance gap among models becomes small. For example, as shown in Figure~\ref{benchmark}, the AUC of most CTR models ranges from 81.35 to 81.50. Moreover, some studies demonstrate that DNN, in practice, struggle to learn multiplicative feature interactions within a limited representation space ~\cite{neuralvsmf, dcnv2} and often occupy a large proportion of network parameters in large-scale production data~\cite{dcnv2}. Taken together, these findings suggest a promising direction for overcoming the performance bottleneck: \textit{Can we remove the dependence on DNN in CTR modeling paradigm and instead employ a truly "deep cross" network to capture extremely high-order explicit feature interactions?}

To answer this question, we revisit the DCNv2 model~\cite{dcnv2} and decompose the modeling process of CrossNetv2 into an aggregation step and an interaction step. Furthermore, we conduct both theoretical and empirical analyses of these two steps. Our results reveal that approximately half of the weight parameters in the aggregation step are relatively redundant, while in the interaction step, the model only considers the interactions between the aggregated feature information and first-order features, which leads to feature interactions with an inefficient linear growth.

To address these limitations and achieve a truly "deep cross" network, this paper proposes a novel explicit feature interactions model for CTR prediction, called the \textbf{Fusing Cross Network (FCN)}. Specifically, FCN consists of two complementary sub-networks: (1) the \textbf{Exponential Cross Network (ECN)}, which aims to capture exponentially growing extremely high-order feature interactions; and (2) the \textbf{Linear Cross Network (LCN)}, which focuses on capturing linearly growing low-order feature interactions. Besides, we introduce a \textbf{Low-cost Aggregation} method to alleviate computational redundancy in the aggregation step, reducing the number of parameters by nearly 50\% and inference latency by about 23\%, with negligible impact on model performance. Moreover, we propose a simple yet effective loss function, named \textbf{Tri-BCE}, which provides appropriate supervision signals for different sub-networks. The core contributions of this paper are summarized as follows:
\begin{itemize}[leftmargin=*]
\item To the best of our knowledge, this is the first work to achieve surprising performance using only explicit feature interaction modeling without integrating DNN.
\item We propose a novel network, called ECN, which captures extremely high-order explicit feature interactions whose order increases exponentially with network depth. To reduce computational redundancy, we introduce a low-cost aggregation method. Furthermore, we design a simple yet effective loss function, Tri-BCE, to provide tailored supervision signals for different sub-networks.
\item We propose a novel CTR model and design two fusion architectures, FCN$_p$ and FCN$_{sp}$, which can adapt to various data distributions by capturing feature interactions of suitable orders.
\item Comprehensive experiments on six benchmark datasets and 16 baselines demonstrate the effectiveness and efficiency of FCN. Based on our experimental results, our models achieve 1st rankings on multiple CTR prediction benchmarks.
\end{itemize}

\section{Related Work}
Effectively capturing feature interactions has always been one of the key methods for improving CTR prediction, thus receiving extensive research attention \cite{EulerNet, GDCN, FINAL}. Traditional methods include LR \cite{LR}, which captures first-order feature interactions, and FM \cite{FM} and its derivatives \cite{FMFM, AFM, FwFM}, which capture second-order feature interactions. With the rise of deep learning, several models attempt to use DNN to capture high-order feature interactions (e.g., PNN \cite{pnn1}, Wide \& Deep \cite{widedeep}, DeepFM \cite{deepfm}, DCNv1 \cite{dcn}, DCNv2 \cite{dcnv2}, SimCEN \cite{SimCEN}, RFM \cite{RFM}, and DIN \cite{DIN}), achieving better performance. Among these, the DCN series models are widely recognized for their effective trade-off between efficiency and performance, gaining significant attention from both academia and industry \cite{dcn, dcnv2, EDCN, GDCN, Xcrossnet, xdeepfm, OptFusion}. Most subsequent deep CTR models follow the paradigm established by DCN, integrating explicit and implicit feature interactions. 

Explicit feature interactions are often modeled directly through hierarchical structures, such as the Cross Layer in the DCN \cite{dcn}, the Graph Layer in FiGNN \cite{fignn}, and the Interacting Layer in AutoInt \cite{autoint}. These methods ensure partial interpretability while allowing the capture of finite-order feature interactions. On the other hand, some studies attempt to integrate implicit feature interactions by designing different structures. These structures mainly include stacked structures \cite{Xcrossnet, pnn1, pnn2}, parallel structures \cite{FINAL, GDCN, CETN}, and alternate structures \cite{SimCEN, FINAL}. The introduction of these structures not only enhances the expressive power of the models but also captures high-order feature interactions through DNN, leading to significant performance improvements in practical applications.

However, as the performance of explicit feature interactions is generally weaker than that of implicit feature interactions \cite{finalmlp}, several models attempt to abandon standalone explicit interaction methods and instead integrate multiplicative operations into DNN. MaskNet \cite{masknet} introduces multiplicative operations block by block, while GateNet \cite{gatenet}, PEPNet \cite{PEPNET}, FINAL \cite{FINAL}, and QNN-$\alpha$ \cite{QNN} introduce them layer by layer to achieve higher performance. Moreover, implicit large language model augmentation methods \cite{HiT-LBM, TrackRec, personax} have also contributed to advancements in the CTR prediction task. Nevertheless, most models struggle to explicitly capture extremely high-order feature interactions and fail to provide appropriate supervision signals for different sub-networks. This paper aims to address these limitations through our proposed methods.

\section{Preliminary}
\subsection{CTR Prediction} It is typically considered a binary classification task that utilizes user profiles, item attributes, and context as features to predict the probability of a user clicking on an item \cite{openbenchmark, autoint}. The composition of these three types of features is as follows:
\begin{itemize}[leftmargin=*]
\item  \emph{User profiles} ($x_U$): age, gender, occupation, etc.
\item \emph{Item attributes} ($x_I$): brand, price, category, etc.
\item \emph{Context} ($x_C$): timestamp, device, position, etc.
\end{itemize}
Further, we can define a CTR input sample in the tuple data format: $X = \{x_U, x_I, x_C\}$. $y \in \{0, 1\}$ is an true label for user click behavior:
\begin{equation}
    y= \begin{cases}1, & \text{user} \text { has clicked } \text{item}, \\ 0, & \text {otherwise, }\end{cases}
\end{equation}
where $y=1$ represents a positive sample and $y=0$ represents a negative sample. A CTR prediction model aims to predict $y$ and rank items based on the predicted probabilities $\hat{y}$.

\subsection{Embedding Layer}
The input feature $X$ of the CTR prediction task, which is multi-field categorical data and is represented using one-hot encoding. Most CTR prediction models \cite{autoint, CETN, adagin} utilize an embedding layer to transform them into low-dimensional dense vectors: $\boldsymbol{e}_i=\textit{E}_i x_i$, where $\textit{E}_i \in \mathbb{R}^{d \times s_i}$ and $s_i$ separately indicate the feature field embedding matrix and the vocabulary size for the $i$-th field, $d$ represents the embedding dimension. Finally, we concatenate these dense vectors to obtain the input $\boldsymbol{x}_1 = \left[\boldsymbol{e}_{(1,1)}, \boldsymbol{e}_{(1,2)}, \cdots, \boldsymbol{e}_{(1,f)}\right] \in \mathbb{R}^D$ of the feature interaction layer, where $D=\sum_{i=1}^f d$, $f$ denotes the number of fields, $\boldsymbol{e}_{(l,i)}  \in \mathbb{R}^{d}$ denotes the $l$-th order feature of the $i$-th feature field, and $\boldsymbol{x}_1$ denotes the original \textbf{first-order features.}\footnote{ In this paper, to ensure that the number of network layers aligns with the order of feature interactions, we take the input $\boldsymbol{x}_1$ as the first-order feature.}

\section{Revisiting Feature Interactions in DCNv2}
\label{Revisiting_DCNv2}
DCNv2~\cite{dcnv2} is a widely recognized modeling paradigm in the CTR prediction task. To better understand how DCNv2 works, we conduct an in-depth analysis of its mechanisms.
\subsection{Implicit Feature Interaction} 
It aims to automatically learn complex non-manually defined data patterns and high-order feature interactions using DNN \cite{xdeepfm, dcnv2}. Formally, for a given input feature $\boldsymbol{x}^1$, the implicit feature interaction process can be expressed as:
\begin{equation}
    \boldsymbol{h}_{k+1}=\sigma\left(\mathbf{W}_{k} \boldsymbol{h}_{k}+\boldsymbol{b}_{k}\right),  \ \ k=1,2,\dots,K,
\end{equation}
where $\boldsymbol{h}_1 = \boldsymbol{x}_1$, $\boldsymbol{h}_{k+1}$ denotes the output of the $k$-th layer, and $\sigma$ represents the activation function. Compared to explicit feature interaction, implicit feature interaction does not have a concrete interaction form, but it learns the inherent data distribution \cite{xdeepfm, dcnv2}. It is highly efficient and performs well~\cite{openbenchmark}, but it has difficulty learning multiplicative feature interactions~\cite{neuralvsmf, dcnv2}.

\subsection{Explicit Feature Interaction} 
\subsubsection{\textbf{Re-Analysis for CrossNetv2}}
Explicit feature interaction seeks to directly capture the combinations and relationships among input features by employing predefined multiplicative interaction functions with controllable order. A popular method for explicit feature interaction is \textbf{CrossNetv2} \cite{dcnv2}, which is described as follows:
\begin{equation}
\label{CrossNetv2}
    \boldsymbol{x}_{l+1} = \boldsymbol{x}_1 \odot \left(\mathbf{W}_l \boldsymbol{x}_l+\boldsymbol{b}_l\right) + \boldsymbol{x}_l,  \ \ l=1,2,\dots,L,
\end{equation}
where $\boldsymbol{x}_l \in \mathbb{R}^{D}$ denotes the features of $l$-th order, $\odot$ is the Hadamard Product, and $\mathbf{W}_l \in \mathbb{R}^{D \times D}$ and $\boldsymbol{b}_l \in \mathbb{R}^{D}$ are the learnable weight matrix and bias vector at $l$-th layer. This method uses the Hadamard Product to interact feature $\boldsymbol{x}_{l}$ with \textbf{anchor features}\footnote{To more intuitively illustrate the similarities and differences between linear and exponential feature interactions, we refer to the feature that is not aggregated by the weight matrix as the anchor feature.} $\boldsymbol{x}_1$ to generate the $(l+1)$-th order feature $\boldsymbol{x}_{l+1}$. To more intuitively illustrate how Eq. (\ref{CrossNetv2}) performs feature interaction, we decompose it into aggregation and interaction steps, and rewrite it as follows (ignoring the bias and residual terms):
\begin{equation}
\label{aggregation_crossnetv2}
\begin{aligned}
&\text{Aggregation:}\ \  \boldsymbol{c}_l = \mathbf{W}_l \boldsymbol{x}_l = \left[\begin{array}{ccc}
W_{(l,1,1)} & \cdots & W_{(l,1, f)} \\
\vdots & \ddots & \vdots \\
W_{(l,f, 1)} & \cdots & W_{(l,f, f)}
\end{array}\right] \left[\begin{array}{c}
\boldsymbol{e}_{(l,1)} \\
\boldsymbol{e}_{(l,2)} \\
\vdots \\
\boldsymbol{e}_{(l,f)}
\end{array}\right] \\ 
& = \left[
\sum_{i=1}^{f}W_{(l,1,i)}\boldsymbol{e}_{(l,i)}, \sum_{i=1}^{f}W_{(l,2,i)}\boldsymbol{e}_{(l,i)}, \dots, \sum_{i=1}^{f}W_{(l,f,i)}\boldsymbol{e}_{(l,i)}\right]^\top,
\end{aligned}
\end{equation}
\begin{equation}
\label{interaction_crossnetv2}
\begin{aligned}
& \text{Interaction:}\ \  \boldsymbol{x}_{l+1} = \boldsymbol{x}_1 \odot \boldsymbol{c}_l = \left[\begin{array}{c}
\boldsymbol{e}_{(1,1)} \odot \sum_{i=1}^{f}W_{\tiny (l,1,i)}\boldsymbol{e}_{(l,i)} \\
\boldsymbol{e}_{(1,2)} \odot \sum_{i=1}^{f}W_{(l,2,i)}\boldsymbol{e}_{(l,i)} \\
\vdots \\
\boldsymbol{e}_{(1,f)} \odot \sum_{i=1}^{f}W_{(l,f,i)}\boldsymbol{e}_{(l,i)} 
\end{array}\right] 
\end{aligned}
\end{equation}
where $W_{(l,i,j)} \in \mathbb{R}^{d \times d}$ represents the importance of interaction between the $i$-th and $j$-th feature fields at the $l$-th layer, and $\boldsymbol{c}_l$ aggregates all $l$-order feature at the $l$-th layer. From Eqs.~(\ref{aggregation_crossnetv2}) and (\ref{interaction_crossnetv2}), we observe that the seemingly simple operation $\boldsymbol{x}_1 \odot \boldsymbol{c}_l$ actually accomplishes two steps:
\begin{itemize}[leftmargin=*]
\item \textbf{Aggregation Step}: It aggregates information from all feature fields for each feature field.
\item \textbf{Interaction Step}: It interacts the aggregated information with \textit{first-order features} to generate higher-order features.
\end{itemize}

This simple two-step approach implements explicit feature interactions with linear growth. DCNv2 combines this approach with implicit feature interactions, thereby further enhancing the model's ability to capture complex feature relationships. However, CrossNetv2 encounters two issues that require further investigation.

\begin{figure*}[t]
    \subfloat[A linear growth feature interaction]{
        \centering
        \includegraphics[width=1\linewidth]{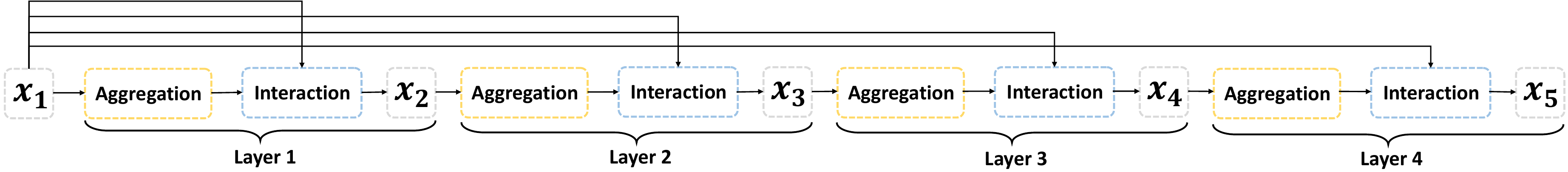}
    } \\
    \subfloat[A exponential  growth feature interaction]{
        \centering
        \includegraphics[width=1\linewidth]{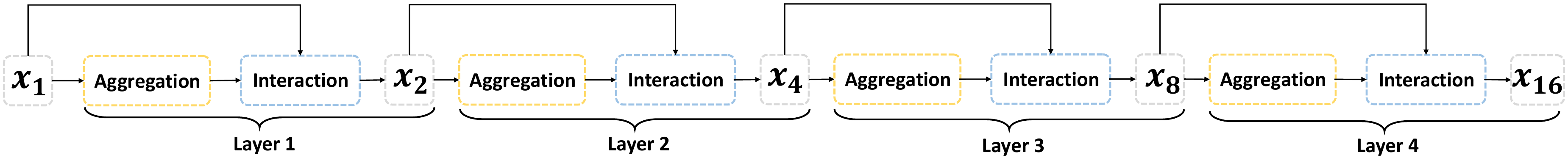}
    } 
    \captionsetup{justification=raggedright}
    \caption{A comparison between feature interaction methods with linear and exponential growth.}
    \label{linear_vs_exp}
\end{figure*}

\begin{figure}[t]
  \centering
  \includegraphics[width=0.5\textwidth]{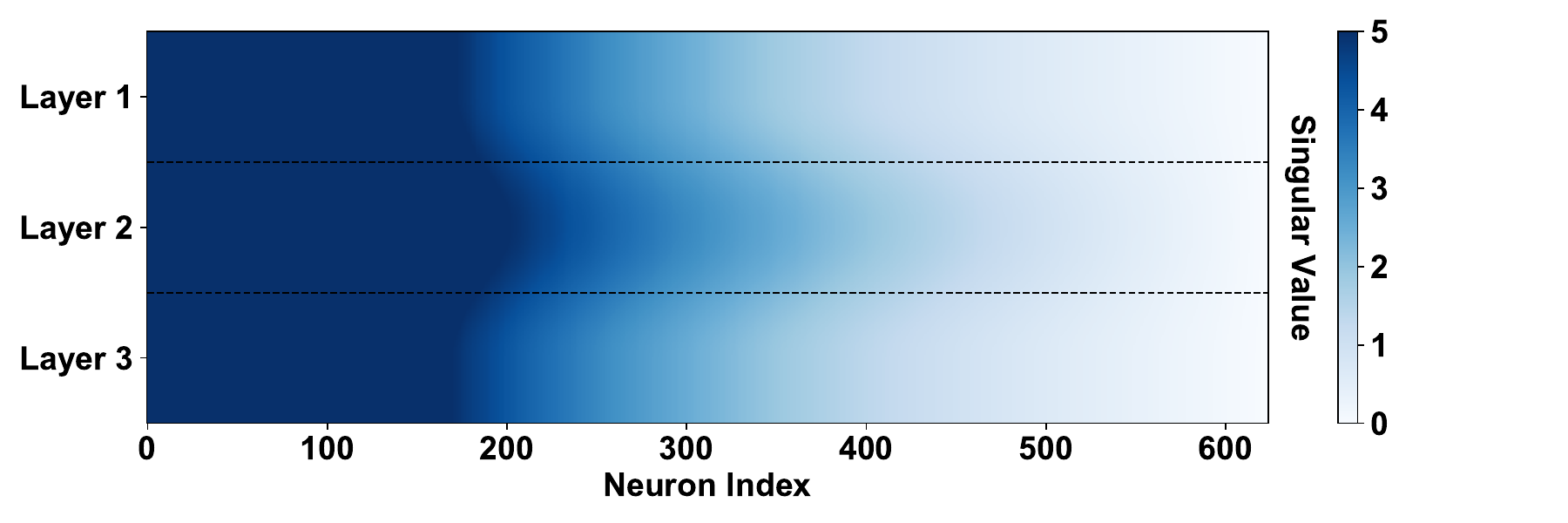}
  \captionsetup{justification=raggedright}
  \caption{The singular value distribution \cite{SVD} of the weight matrix $\mathbf{W}_l$ in each layer of CrossNetv2 on Criteo dataset.}
  \label{CrossNetv2_low_rank}
  \vspace{-2em}
\end{figure}

\subsubsection{\textbf{Computational Redundancy in the Aggregation Step}}
\label{section_Redundancy}
To investigate whether the weight matrix $\mathbf{W}_l$ in the aggregation step fully utilizes the parameter space, we perform singular value decomposition (SVD)~\cite{SVD} and analyze the distribution of singular values to determine whether computational redundancy exists. Specifically, for a parameter matrix $\mathbf{W}$, we perform SVD as $\mathbf{W} = \boldsymbol{U} \boldsymbol{\Sigma} \boldsymbol{V}^\top$, where $\boldsymbol{\Sigma}$ denotes the incrementally ordered singular value distribution of $\mathbf{W}$. A smaller singular value in a particular dimension indicates lower utilization and more severe redundancy~\cite{feature_Collapse, Dimensional_Collapse}. The experimental results are presented in Figure \ref{CrossNetv2_low_rank}. We observe that nearly half of the singular values of the weight matrix $\mathbf{W}_l$ in CrossNetv2 are relatively small, indicating redundancy in the parameter space. This finding motivates us to explore a more lightweight method for measuring the importance of interactions between feature fields.

\subsubsection{\textbf{Inefficient Feature Interaction in the Interaction Step}}
To provide a more intuitive explanation of why CrossNetv2 implements feature interactions with linear growth, we visualize its interaction process in Figure~\ref{linear_vs_exp} (a). Consistent with Eq.~(\ref{interaction_crossnetv2}), each layer of CrossNetv2 first aggregates information from all feature fields and then interacts with $\boldsymbol{x}_1$ to generate the $\boldsymbol{x}_{l+1}$ feature. This method, which fixes the interaction to $\boldsymbol{x}_1$ after aggregation, limits the efficiency of feature interactions.

To address the inefficiency in feature interaction modeling, a simple yet effective method is to replace the anchor feature $\boldsymbol{x}_{1}$ with $\boldsymbol{x}_{2^{l-1}}$, as illustrated in Figure~\ref{linear_vs_exp} (b). By interacting the aggregated information with $\boldsymbol{x}_{2^{l-1}}$ at each layer, we achieve exponentially growing feature interactions. As shown in Figure~\ref{linear_vs_exp}, when both methods are stacked for four layers, the linear feature interaction can only model up to $\boldsymbol{x}_5$, whereas the exponential feature interaction can model up to $\boldsymbol{x}_{16}$.


\section{Methodology}
 Based on the findings presented in Section~\ref{Revisiting_DCNv2}, we propose the FCN model, shown in Figure \ref{FCN}, which integrates different explicit feature interaction sub-networks: LCN and ECN, enabling it to simultaneously explicitly capture both low-order and high-order feature interactions. 

\subsection{Fusing Cross Network}
Previous deep CTR models~\cite{dcnv2, xdeepfm, deepfm, GDCN} typically employ explicit interaction methods to capture low-order feature interactions and rely on DNN to implicitly capture high-order feature interactions. However, the former is limited to capturing low-order interactions due to complexity constraints and generally exhibits lower performance~\cite{finalmlp, FINAL, openbenchmark}, while the latter struggles to learn multiplicative feature interactions within a limited representation space~\cite{neuralvsmf, dcnv2}. Therefore, we aim to achieve a truly "deep cross" network that captures explicit high-order feature interactions, thereby eliminating the dependence of CTR models on DNN.

\subsubsection{\textbf{Low-cost Aggregation (LCA)}} 
\begin{figure}[t]
  \centering
  \includegraphics[width=0.48\textwidth]{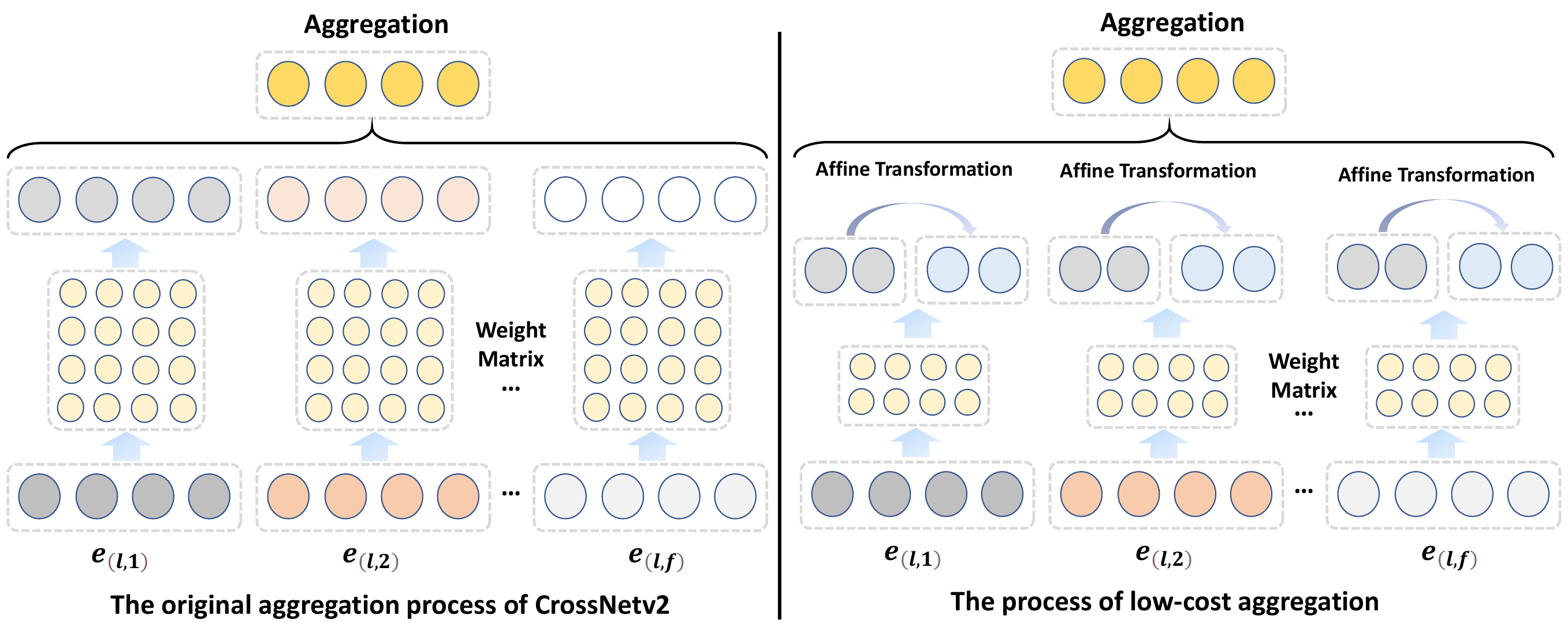}
  \captionsetup{justification=raggedright}
  \caption{A comparison between the original aggregation process and the low-cost aggregation process.}
  \label{LCA_figure}
\end{figure}

As presented in Section~\ref{section_Redundancy}, nearly half of the parameter space of the weight matrix $\mathbf{W}_l$ in the cross network remains underutilized. Therefore, we directly reduce the size of the weight matrix by half and introduce a low-cost affine transformation to maintain consistency with the original output dimension. For clarity, we provide the visualization shown in Figure~\ref{LCA_figure}. Formally, we define this process as follows:
\begin{equation}  
\label{LCA_equ}
\begin{aligned}
\text{LCA}\left(\boldsymbol{x}_l\right) &= \left[c_{(l,1)} \Vert c^{\prime}_{(l,1)}, c_{(l,2)} \Vert c^{\prime}_{(l,2)}, \cdots, c_{(l,f)} \Vert c^{\prime}_{(l,f)}\right], \\
c_{(l,i)} &= \sum_{j=1}^{f}\left(W_{(l,i,j)} \boldsymbol{e}_{(l,j)}+ b_{(l,i,j)}\right), \ \ l=1,2,\dots,L,\\
c^{\prime}_{(l,i)} &= \gamma_{(l,i)} \odot c_{(l,i)} + \beta_{(l,i)},
\end{aligned}
\end{equation}
where $W_{(l,i,j)} \in \mathbb{R}^{\frac{d}{2} \times d}$ denotes the importance of interaction between the $i$-th and $j$-th feature fields at the $l$-th layer, $c_{(l,i)}$ is the aggregation vector of all features fields for the $i$-th feature field, $\gamma_{(l,i)}, \beta_{(l,i)} \in \mathbb{R}^{\frac{d}{2}}$ are the affine parameters for the $i$-th feature field, and $\Vert$ denotes the concatenation operation. LCA employs low-cost affine transformations to halve the parameter space required for the aggregation step, thereby enhancing the efficiency of feature information aggregation.

\begin{figure}[t]
  \centering
  \includegraphics[width=0.45\textwidth]{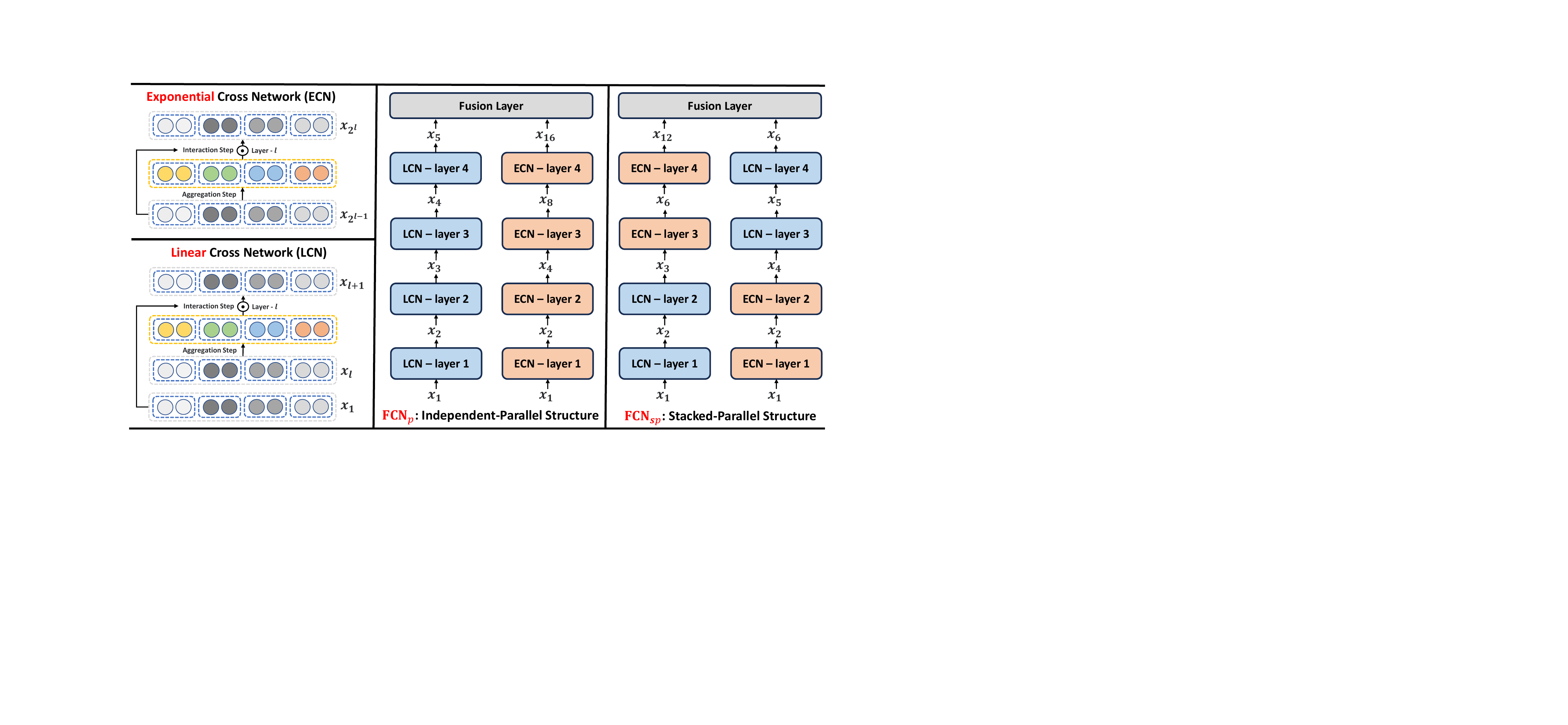}
  \captionsetup{justification=raggedright}
  \caption{An example of the modeling processes of FCN$_p$ and FCN$_{sp}$.}
  \label{FCN}
\end{figure}

\subsubsection{\textbf{Linear Cross Network (LCN)}} LCN adopts the same idea as CrossNetv2. It is utilized to capture low-order explicit feature interactions with linear growth. Its recursive formulation is:
\begin{equation}  
\begin{aligned}
\boldsymbol{x}_{l+1} &= \boldsymbol{x}_1 \odot \text{LCA}\left(\boldsymbol{x}_l\right) + \boldsymbol{x}_l,  \ \ l=1,2,\dots,L,\\
\end{aligned}
\end{equation}
where the anchor features are $\boldsymbol{x}_1$, and the aggregated features $\text{LCA}\left(\boldsymbol{x}_l\right)$ interact with $\boldsymbol{x}_1$ to generate the $(l+1)$-th order feature.

\subsubsection{\textbf{Exponential Cross Network (ECN)}} 
\begin{figure}[t]
  \centering
  \includegraphics[width=0.45\textwidth]{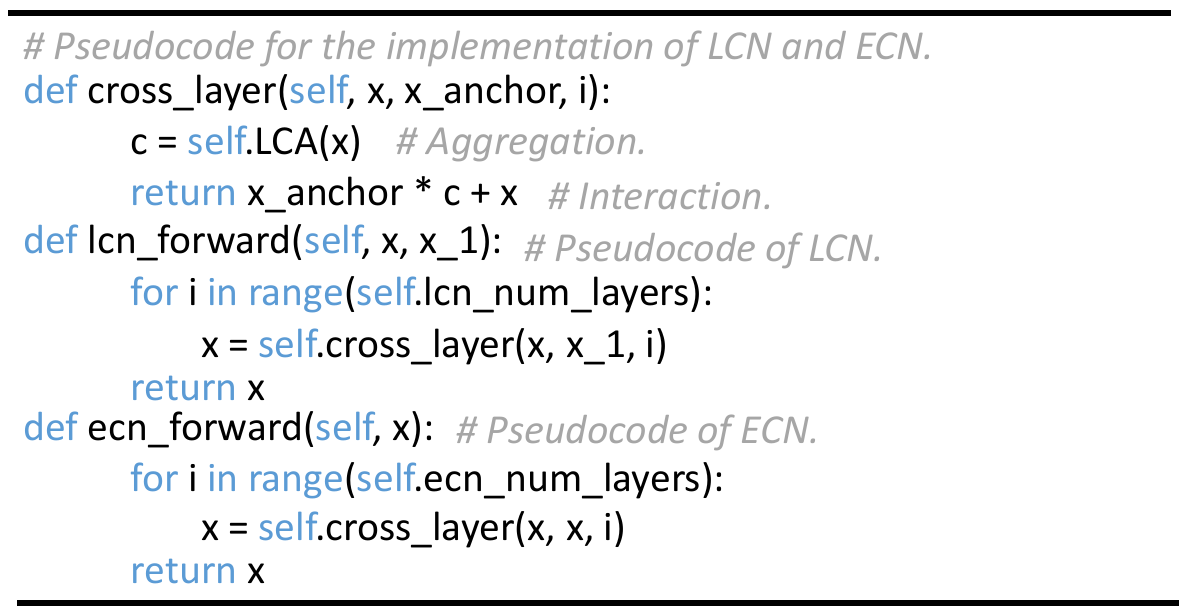}
  \captionsetup{justification=raggedright}
  \caption{Pseudocode for LCN and ECN.}
  \label{LCN_ECN_figure}
\end{figure}

As the core idea of FCN, it is used to capture \textit{extremely high-order} explicit feature interactions with exponential growth. Its recursive formula is:
\begin{equation}
\label{ECN_Equs}
\begin{aligned}
\boldsymbol{x}_{{2^{l}}} &= \boldsymbol{x}_{2^{l-1}} \odot \text{LCA}\left(\boldsymbol{x}_{2^{l-1}}\right) + \boldsymbol{x}_{2^{l-1}},  \ \ l=1,2,\dots,L,\\
\end{aligned}
\end{equation}
where $\boldsymbol{x}_{2^l} \in \mathbb{R}^D$ represents the $2^l$-th order feature. Unlike LCN, ECN modifies the anchor features from $\boldsymbol{x}_1$ to $\boldsymbol{x}_{2^{l-1}}$, thereby achieving high-order feature interactions within a limited number of layers. To more clearly illustrate the similarities and differences between LCN and ECN, we provide pseudocode in Figure~\ref{LCN_ECN_figure}. Combining the results in Figure \ref{benchmark}, we observe that ECN achieves SOTA performance by modifying only a single variable based on LCN.

\subsubsection{\textbf{Fusing LCN and ECN}} Most previous CTR models \cite{FINAL, finalmlp, dcnv2} attempt to model explicit and implicit feature interactions, which essentially means capturing both low-order and high-order feature interactions. Our FCN achieves this by fusing LCN and ECN, avoiding the use of DNN. Specifically, we propose two fusion architectures:
\begin{itemize}[leftmargin=*]
\item \textbf{FCN$_p$: Independent-Parallel architecture}: It allows LCN and ECN to process input features separately.
\item \textbf{FCN$_{sp}$: Stacked-Parallel architecture}: It sequentially stacks one network on top of the other.
\end{itemize}
These two fusion architectures are illustrated in Figure~\ref{FCN}. We observe that FCN$_p$ captures the features [$\boldsymbol{x}_1, \boldsymbol{x}_2, \boldsymbol{x}_3, \boldsymbol{x}_4, \boldsymbol{x}_5, \boldsymbol{x}_8, \boldsymbol{x}_{16}$], while FCN$_{sp}$ captures the features [$\boldsymbol{x}_1, \boldsymbol{x}_2, \boldsymbol{x}_3, \boldsymbol{x}_4, \boldsymbol{x}_5, \boldsymbol{x}_6, \boldsymbol{x}_{12}$]. This indicates that the two architectures focus on different orders of feature interaction modeling. In practice, the choice of fusion architecture can be adjusted according to the order of feature interactions required by the data distribution.

Finally, we use a simple linear transformation to convert the output representation of the FCN into the final prediction\footnote{Other advanced ensemble methods, such as DHEN~\cite{DHEN} and HMoE~\cite{HMoE}, can also be applied here.}. Taking FCN$_p$ as an example, the fusion layer is formalized as follows:
\begin{equation}  
\begin{aligned}
&\hat{y} = \texttt{Mean}(\hat{y}_{\scriptscriptstyle exp}, \hat{y}_{\scriptscriptstyle lin}), \\
\hat{y}_{\scriptscriptstyle exp} = \sigma(\mathbf{W}_{\scriptscriptstyle exp}\boldsymbol{x}_{{2^{L}}} + &\  \boldsymbol{b}_{\scriptscriptstyle exp}),\ \ 
\hat{y}_{\scriptscriptstyle lin} = \sigma(\mathbf{W}_{\scriptscriptstyle lin}\boldsymbol{x}_{L+1} + \boldsymbol{b}_{\scriptscriptstyle lin}),
\end{aligned}
\end{equation}
where $\mathbf{W}_{\scriptscriptstyle exp}$ and $\mathbf{W}_{\scriptscriptstyle lin} \in \mathbb{R}^{1 \times D}$ represent learnable weights, $\boldsymbol{b}_{\scriptscriptstyle exp}$ and $\boldsymbol{b}_{\scriptscriptstyle lin}$ are biases, \texttt{Mean} denotes the mean fusion, $\hat{y}_{\scriptscriptstyle exp}, \hat{y}_{\scriptscriptstyle lin}$, $\hat{y}$ represent the prediction results of ECN, LCN, and FCN, respectively, and $L$ denotes the last number of layers. 

\subsubsection{\textbf{Tri-BCE Loss}}
\begin{figure}[t]
  \centering
  \includegraphics[width=0.45\textwidth]{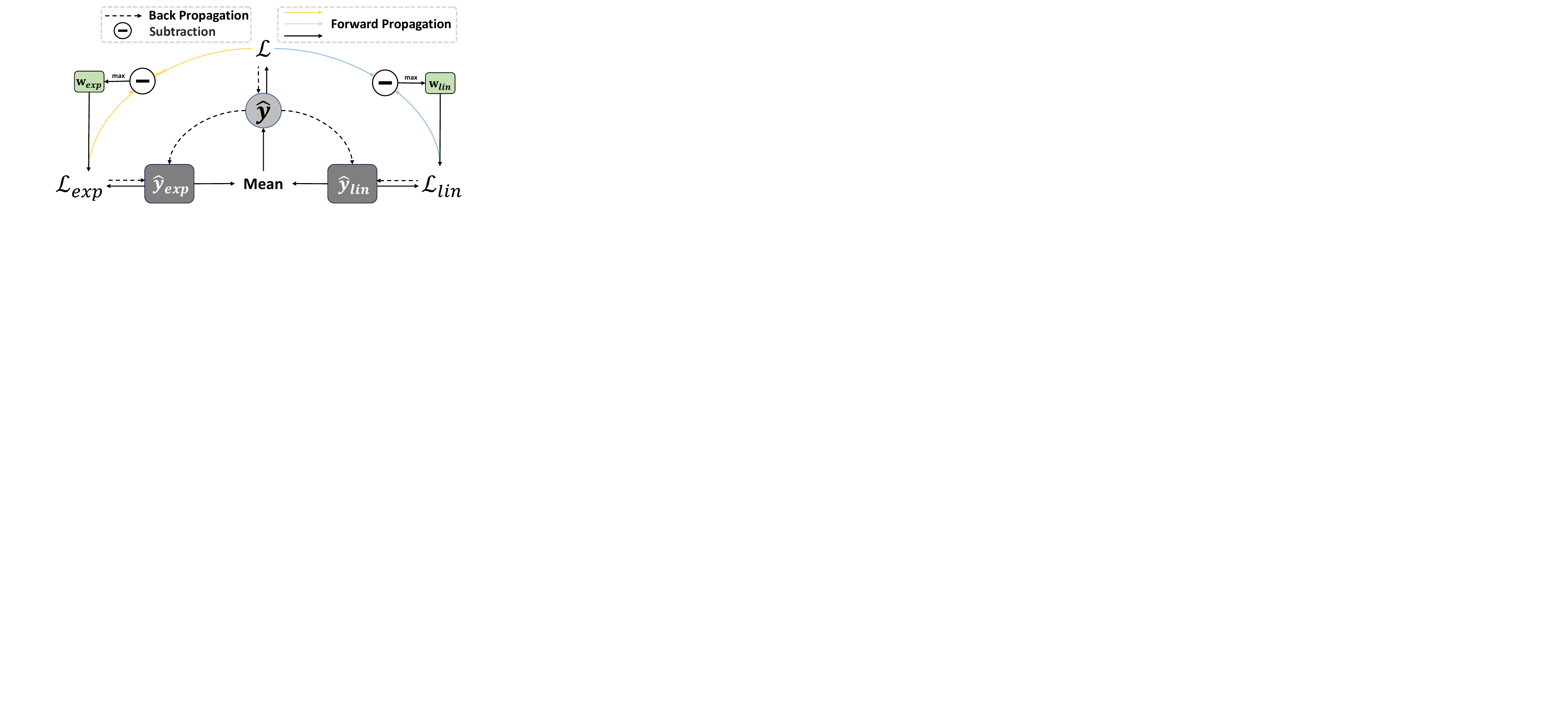}
   
  \captionsetup{justification=raggedright}
  \caption{\textbf{The workflow for the Tri-BCE loss.}}
  \label{fusion}
   
\end{figure}

In most CTR prediction models \cite{finalmlp, GDCN, dcnv2}, the loss is typically computed solely based on the final prediction $\hat{y}$, overlooking the distinct supervision signals required by individual sub-networks. To address this, we propose Tri-BCE, which provides tailored supervision signals for different sub-networks. The calculation process and balancing method of the multi-loss are illustrated in Figure \ref{fusion}. We use the widely adopted binary cross-entropy (BCE) loss \cite{FINAL, finalmlp} (i.e., Logloss) as both the primary and auxiliary loss for FCN:
\begin{equation}  
\label{loss}
\begin{aligned}
\mathcal{L}&=-\frac{1}{N} \sum_{i=1}^N\left(y_i \log \left(\hat{y}_i\right)+\left(1-y_i\right) \log \left(1-\hat{y}_i\right)\right), \\
\mathcal{L}_{\scriptscriptstyle exp} &=-\frac{1}{N} \sum_{i=1}^N\left(y_i \log \left(\hat{y}_{\scriptscriptstyle exp,i}\right)+\left(1-y_i\right) \log \left(1-\hat{y}_{\scriptscriptstyle exp,i}\right)\right), \\
\mathcal{L}_{\scriptscriptstyle lin} &=-\frac{1}{N} \sum_{i=1}^N\left(y_i \log \left(\hat{y}_{\scriptscriptstyle lin,i}\right)+\left(1-y_i\right) \log \left(1-\hat{y}_{\scriptscriptstyle lin,i}\right)\right), \\
\end{aligned}
\end{equation}
where $y$ denotes the true labels, $N$ denotes the batch size, $\mathcal{L}_{\scriptscriptstyle exp}$ and $\mathcal{L}_{\scriptscriptstyle lin}$ represent the auxiliary losses for the prediction results of ECN and LCN, respectively, and $\mathcal{L}$ represents the primary loss. To provide each sub-network with suitable supervision signals, we assign them adaptive weights, $\mathbf{w}_{\scriptscriptstyle exp} = \texttt{max}(0, \mathcal{L}_{\scriptscriptstyle exp} - \mathcal{L})$ and $\mathbf{w}_{\scriptscriptstyle lin} = \texttt{max}(0, \mathcal{L}_{\scriptscriptstyle lin} - \mathcal{L})$, and jointly train them to achieve Tri-BCE:
\begin{equation}  
\begin{aligned}
\mathcal{L}_{\text{Tri}} &= \mathcal{L} + \mathbf{w}_{\scriptscriptstyle exp} \cdot \mathcal{L}_{\scriptscriptstyle exp} + \mathbf{w}_{\scriptscriptstyle lin} \cdot \mathcal{L}_{\scriptscriptstyle lin},
\end{aligned}
\end{equation}
as demonstrated by \cite{TF4CTR}, providing a single supervision signal to sub-networks is often suboptimal. Our proposed Tri-BCE loss helps sub-networks learn better parameters by providing adaptive weights that change throughout the learning process. Theoretically, we can derive the gradients obtained by $\hat{y}_{\scriptscriptstyle exp}$:
\begin{equation}
\begin{aligned}
\nabla_{(\hat{y}^{+}_{\scriptscriptstyle exp})} \mathcal{L}_{\text{Tri}} & = -\frac{1}{N} \cdot \frac{\partial \left(\log \hat{y}^+ + \mathbf{w}_{\scriptscriptstyle exp} \log\hat{y}^{+}_{\scriptscriptstyle exp}\right)}{\partial \hat{y}^{+}_{\scriptscriptstyle exp}} \\
&= -\frac{1}{N}\left(\frac{1}{2\hat{y}^+} + \frac{\mathbf{w}_{\scriptscriptstyle exp}}{\hat{y}^{+}_{\scriptscriptstyle exp}}\right), \\
\nabla_{(\hat{y}^{-}_{\scriptscriptstyle exp})} \mathcal{L}_{\text{Tri}} & =  -\frac{1}{N} \cdot \frac{\partial \left(\log (1 - \hat{y}^-) + \mathbf{w}_{\scriptscriptstyle exp} \log(1 - \hat{y}^{-}_{\scriptscriptstyle exp})\right)}{\partial \hat{y}^{-}_{\scriptscriptstyle exp}}  \\
&= \frac{1}{N}\left(\frac{1}{2(1-\hat{y}^-)} + \frac{\mathbf{w}_{\scriptscriptstyle exp}}{1-\hat{y}^{-}_{\scriptscriptstyle exp}}\right),
\end{aligned}
\end{equation}
where $\nabla_{(\hat{y}^{+}_{\scriptscriptstyle exp})}$ and $\nabla_{(\hat{y}^{-}_{\scriptscriptstyle exp})}$ represent the gradients received by $\hat{y}_{\scriptscriptstyle exp}$ for positive and negative samples, respectively. Similarly, the gradient signals received by $\hat{y}_{\scriptscriptstyle lin}$ are consistent with those of $\hat{y}_{\scriptscriptstyle exp}$, so we do not elaborate further. It can be observed that $\hat{y}_{\scriptscriptstyle exp}$ and $\hat{y}_{\scriptscriptstyle lin}$ both have the same gradient terms, $\frac{1}{2\hat{y}^+}$ and $\frac{1}{2(1-\hat{y}^-)}$, indicating that training both sub-networks with a single loss provides identical supervision signals for both, making it difficult for the two sub-networks to learn and specialize in different types of feature interactions. However, our Tri-BCE loss additionally provides dynamically adjusted gradient terms based on $\mathbf{w}_{\scriptscriptstyle exp}$ and $\mathbf{w}_{\scriptscriptstyle lin}$, ensuring that the sub-networks are directly influenced by the true labels $y$ and adaptively adjust their weights according to the difference between the primary and auxiliary losses.

\subsection{Theoretical Analysis}
\subsubsection{\textbf{ECN is superior to LCN}}
To further theoretically clarify the differences between ECN and LCN, we rewrite Eq. (\ref{ECN_Equs}) by following the aggregation and interaction steps:
\begin{equation}
\label{interaction_ecn}
\begin{aligned}
\boldsymbol{x}_{2^{l}} &= \boldsymbol{x}_{2^{l-1}} \odot \text{LCA}\left(\boldsymbol{x}_{2^{l-1}}\right), \\ 
&=\left[\begin{array}{c}
\boldsymbol{e}_{(2^{l-1},1)} \odot \left[c_{(2^{l-1},1)} \Vert c^{\prime}_{(2^{l-1},1)}\right] \\
\boldsymbol{e}_{(2^{l-1},2)} \odot \left[c_{(2^{l-1},2)} \Vert c^{\prime}_{(2^{l-1},2)}\right] \\
\vdots \\
\boldsymbol{e}_{(2^{l-1},f)} \odot \left[c_{(2^{l-1},f)} \Vert c^{\prime}_{(2^{l-1},f)}\right] \\
\end{array}\right], 
\end{aligned}
\end{equation}
where $\boldsymbol{e}_{(2^{l-1},i)} \in \mathbb{R}^{d}$ denote the $2^{l-1}$-th order features of the $i$-th feature field. By analyzing Eqs. (\ref{interaction_ecn}) and (\ref{LCA_equ}), we observe that, when weight matrices are disregarded, the fully expanded recursive formulation of ECN implements a feature interaction process that can be simplified as:
\begin{equation}
\begin{aligned}
\boldsymbol{e}_{(2^{l},i)}&=\boldsymbol{e}_{(2^{l-1},i)} \odot \sum_{i=1}^{f} \boldsymbol{e}_{(2^{l-1},i)},\\
&=\boldsymbol{e}_{(1,i)} \odot \sum_{i=1}^{f} \boldsymbol{e}_{(1,i)} \odot \cdots \odot \sum_{i=1}^{f} \boldsymbol{e}_{(2^{l-2},i)} \odot \sum_{i=1}^{f} \boldsymbol{e}_{(2^{l-1},i)}.
\end{aligned}
\end{equation}
Meanwhile, the feature interaction process of LCN, consistent with CrossNetv2, is expressed as:
\begin{equation}
\begin{aligned}
\boldsymbol{e}_{(l+1,i)}=\boldsymbol{e}_{(1,i)} \odot \sum_{i=1}^{f} \boldsymbol{e}_{(l,i)}.
\end{aligned}
\end{equation}
Compared to LCN, ECN facilitates a more sophisticated and comprehensive feature interaction. Through multi-layer recursive expansion, ECN captures higher-order feature interactions, significantly enhancing the CrossNet’s expressive capacity.

\subsubsection{\textbf{Complexity Analysis}}
\label{Complexity}
To further compare the efficiency of the DCN series models, we discuss and analyze the time complexity of different models. Let $W_{\Psi}$ denote the predefined number of parameters in the DNN. The definitions of the other variables can be found in the previous sections. For clarity, we further provide a comparison of the magnitudes of different variables in Table \ref{TimeComplexity}. We can derive:

\begin{table}[htbp]
\renewcommand\arraystretch{1.1}
\Huge
\centering
\caption{Comparison of Analytical Time Complexity \\
$s \gg  |W_{\Psi}| > D > f \approx d > L$}
\resizebox{0.9\linewidth}{!}{
\begin{tabular}{c|c|c|c}
\toprule  \textbf { Model } & \textbf { Embedding } & \textbf {Implicit interaction} & \textbf {Explicit interaction} \\ \hline
\text { DCNv1 \cite{dcn}} & \textit{O}($dfs$) & \textit{O}($|W_{\Psi}|$) & \textit{O}($2DL$) \\
 \text { DCNv2 \cite{dcnv2}} & \textit{O}($dfs$) & \textit{O}($|W_{\Psi}|$) & \textit{O}($D^2L$) \\
 \text { EDCN \cite{EDCN}} & \textit{O}($dfs$) & \textit{O}($D^2L$) & \textit{O}($D^2L$) \\
\text { GDCN \cite{GDCN}} & \textit{O}($dfs$) & \textit{O}($|W_{\Psi}|$) & \textit{O}($2D^2L$) \\
 \textbf { ECN } & \textit{O}($dfs$) & - & \textit{O}($D^2L / 2$) \\
 \textbf { FCN$_p$ \& FCN$_{sp}$ } & \textit{O}($dfs$) & - & \textit{O}($D^2L$) \\
\bottomrule
\end{tabular}}
\label{TimeComplexity}
 
\end{table}

\begin{itemize}[leftmargin=*]
\item Except for our proposed ECN and FCN, all other models include implicit interaction to enhance predictive performance, which incurs additional computational costs.
\item In terms of explicit interaction, ECN only has a higher time complexity than DCNv1, and the time complexity of GDCN is four times that of ECN.
\item Our FCN model uses the Tri-BCE loss function, which theoretically has a time complexity for loss computation three times higher than other models. However, in practical training, due to optimizations in parallel computation, its training cost is comparable to some models already deployed in production environments (e.g. FinalMLP\cite{finalmlp}, FINAL\cite{FINAL}) and does not reach the theoretical threefold increase. This is validated in Figure \ref{EfficiencyComparisons}. Moreover, this design has no impact on the inference speed.
\end{itemize}

\section{Experiments}
\subsection{Experiment Setup}
\subsubsection{\textbf{Datasets.}} We evaluate FCN on six CTR prediction datasets: Avazu\footnote{\url{https://www.kaggle.com/c/avazu-ctr-prediction}} \cite{openbenchmark}, Criteo\footnote{ \url{https://www.kaggle.com/c/criteo-display-ad-challenge}} \cite{openbenchmark}, ML-1M\footnote{ \url{https://grouplens.org/datasets/movielens}} \cite{autoint}, KDD12\footnote{\url{https://www.kaggle.com/c/kddcup2012-track2}} \cite{autoint}, iPinYou\footnote{\url{https://contest.ipinyou.com/}} \cite{pnn2}, and KKBox\footnote{\url{https://www.kkbox.com/intl}} \cite{Bars}. Table \ref{dataset}  provides detailed information about these datasets. A more detailed description of these datasets can be found in the given references and links.

\begin{table}[t]
\tiny
\renewcommand\arraystretch{1}
\centering
\caption{Dataset statistics}
\label{dataset}
 
\resizebox{0.8\linewidth}{!}{
\begin{tabular}{cccc} 
\toprule 
\textbf{Dataset} & \textbf{\#Instances} & \textbf{\#Fields} & \textbf{\#Features} \\
\midrule 
\textbf{Avazu}  & 40,428,967  & 24 & 3,750,999 \\
\textbf{Criteo} & 45,840,617 & 39  & 910,747 \\
\textbf{ML-1M} & 739,012  & 7  & 9,751 \\
\textbf{KDD12} & 141,371,038  & 13  & 4,668,096 \\
\textbf{iPinYou} & 19,495,974 & 16  & 665,765\\
\textbf{KKBox} & 7,377,418 & 13  & 91,756\\
\textbf{Industrial} & $\approx$ 10 Billion & $\approx$ 200  & \textbackslash \\
\bottomrule
\end{tabular}}
 
\end{table}

\subsubsection{\textbf{Data Preprocessing.}} We follow the approach outlined in \cite{openbenchmark}. For the Avazu dataset, we transform the timestamp field it contains into three new feature fields: hour, weekday, and weekend. For the Criteo and KDD12 dataset, we discretize the numerical feature fields by rounding down each numeric value $x$ to $\lfloor \log^2(x) \rfloor$ if $x > 2$, and $x = 1$ otherwise. We set a threshold to replace infrequent categorical features with a default "OOV" token. We set the threshold to 10 for Criteo, KKBox, and KDD12, 2 for Avazu and iPinYou, and 1 for the small dataset ML-1M. More specific data processing procedures and results can be found in our open-source run logs\footref{footnote:checkpoint} and configuration files, which we do not elaborate on here.

\begin{table*}[t]
\small
\renewcommand\arraystretch{0.8}
\centering
\caption{Performance comparison of different deep CTR models. "*": Integrating the original model with DNN networks. Meanwhile, we conduct a two-tailed T-test to assess the statistical significance between our models and the best baseline ($\star$: $p$ < 1e-3). \textit{Abs.Imp} represents the absolute performance improvement of FCN over the strongest baseline. Typically, CTR researchers consider an improvement of \textit{0.001 (0.1\%)} in Logloss and AUC to be statistically significant \cite{dcn,EDCN,CL4CTR,openbenchmark}.} 
\label{baselines}
 
\resizebox{\linewidth}{!}{
\begin{tabular}{ccccccccccccc}
\Xhline{1px}
  \multicolumn{1}{c|}{} &
  \multicolumn{2}{c|}{\textbf{Avazu}} &
  \multicolumn{2}{c|}{\textbf{Criteo}} &
  \multicolumn{2}{c|}{\textbf{ML-1M}} &
  \multicolumn{2}{c|}{\textbf{KDD12}} &
  \multicolumn{2}{c|}{\textbf{iPinYou}} &
  \multicolumn{2}{c}{\textbf{KKBox}}\\ \cline{2-13} 
  \multicolumn{1}{c|}{\multirow{-2}{*}{\textbf{Models}}} &
  \multicolumn{1}{c}{Logloss$\downarrow$} &
  \multicolumn{1}{c|}{AUC(\%)$\uparrow$} &
  \multicolumn{1}{c}{Logloss$\downarrow$} &
  \multicolumn{1}{c|}{AUC(\%)$\uparrow$} & 
  \multicolumn{1}{c}{Logloss$\downarrow$} &
  \multicolumn{1}{c|}{AUC(\%)$\uparrow$} &
  \multicolumn{1}{c}{Logloss$\downarrow$} &
  \multicolumn{1}{c|}{AUC(\%)$\uparrow$} &
  \multicolumn{1}{c}{Logloss$\downarrow$} &
  \multicolumn{1}{c|}{AUC(\%)$\uparrow$} &
  \multicolumn{1}{c}{Logloss$\downarrow$} &
  \multicolumn{1}{c}{AUC(\%)$\uparrow$} \\
  \hline
  \multicolumn{1}{c|}{DNN \cite{DNN}} &
  0.3721 &
  \multicolumn{1}{c|}{79.27} &
  0.4380 &
  \multicolumn{1}{c|}{81.40} & 
  0.3100 &
  \multicolumn{1}{c|}{90.30} &
  0.1502 &
  \multicolumn{1}{c|}{80.52} &
   0.005545 &
  \multicolumn{1}{c|}{78.06} &
  0.4811 &
  \multicolumn{1}{c}{85.01} \\
  \multicolumn{1}{c|}{PNN \cite{pnn1}} &
  0.3712 &
  \multicolumn{1}{c|}{79.44} &
  0.4378 &
  \multicolumn{1}{c|}{81.42} &
  0.3070 &
  \multicolumn{1}{c|}{90.42} &
  0.1504 &
  \multicolumn{1}{c|}{80.47} &
  0.005544 &
  \multicolumn{1}{c|}{78.13} &
  0.4793 &
  \multicolumn{1}{c}{85.15} \\
  \multicolumn{1}{c|}{Wide \& Deep \cite{widedeep}} &
  0.3720 &
  \multicolumn{1}{c|}{79.29} &
  0.4376 &
  \multicolumn{1}{c|}{81.42} &
  0.3056 &
  \multicolumn{1}{c|}{90.45} &
  0.1504 &
  \multicolumn{1}{c|}{80.48} &
  0.005542 &
  \multicolumn{1}{c|}{78.09} &
  0.4852 &
  \multicolumn{1}{c}{85.04} \\
  \multicolumn{1}{c|}{DeepFM \cite{deepfm}} &
  0.3719 &
  \multicolumn{1}{c|}{79.30} &
  0.4375 &
  \multicolumn{1}{c|}{81.43} &
  0.3073 &
  \multicolumn{1}{c|}{90.51} &
  0.1501 &
  \multicolumn{1}{c|}{80.60} &
  0.005549 &
  \multicolumn{1}{c|}{77.94} &
  0.4785 &
  \multicolumn{1}{c}{85.31}\\
  \multicolumn{1}{c|}{DCNv1 \cite{dcn}} &
  0.3719 &
  \multicolumn{1}{c|}{79.31} &
  0.4376 &
  \multicolumn{1}{c|}{81.44} &
  0.3156 &
  \multicolumn{1}{c|}{90.38} &
  0.1501 &
  \multicolumn{1}{c|}{80.59} &
  0.005541 &
  \multicolumn{1}{c|}{78.13} &
  \underline{0.4766} &
  \multicolumn{1}{c}{85.31}\\
  \multicolumn{1}{c|}{xDeepFM \cite{xdeepfm}} &
  0.3718 &
  \multicolumn{1}{c|}{79.33} &
  0.4376 &
  \multicolumn{1}{c|}{81.43} &
  0.3054 &
  \multicolumn{1}{c|}{90.47} &
  0.1501 &
  \multicolumn{1}{c|}{80.62} &
  0.005534 &
  \multicolumn{1}{c|}{78.25} &
  0.4772 &
  \multicolumn{1}{c}{\underline{85.35}} \\
  \multicolumn{1}{c|}{AutoInt* \cite{autoint}} &
  0.3746 &
  \multicolumn{1}{c|}{79.02} &
  0.4390 &
  \multicolumn{1}{c|}{81.32} &
  0.3112 &
  \multicolumn{1}{c|}{90.45} &
  0.1502 &
  \multicolumn{1}{c|}{80.57} &
  0.005544 &
  \multicolumn{1}{c|}{78.16} &
  0.4773 &
  \multicolumn{1}{c}{85.34} \\
  \multicolumn{1}{c|}{AFN* \cite{AFN}} &
  0.3726 &
  \multicolumn{1}{c|}{79.29} &
  0.4384 &
  \multicolumn{1}{c|}{81.38} &
  0.3048 &
  \multicolumn{1}{c|}{90.53} &
   0.1499 &
  \multicolumn{1}{c|}{80.70} &
  \underline{0.005539} &
  \multicolumn{1}{c|}{78.17} &
  0.4842 &
  \multicolumn{1}{c}{84.89}\\
  \multicolumn{1}{c|}{DCNv2 \cite{dcnv2}} &
  0.3718 &
  \multicolumn{1}{c|}{79.31} &
  0.4376 &
  \multicolumn{1}{c|}{81.45} &
  0.3098 &
  \multicolumn{1}{c|}{\underline{90.56}} &
  0.1502 &
  \multicolumn{1}{c|}{80.59} &
  \underline{0.005539} &
  \multicolumn{1}{c|}{78.26} &
  0.4787 &
  \multicolumn{1}{c}{85.31}\\
  \multicolumn{1}{c|}{EDCN \cite{EDCN}} &
  0.3716 &
  \multicolumn{1}{c|}{79.35} &
  0.4378 &
  \multicolumn{1}{c|}{81.44} &
  0.3073 &
  \multicolumn{1}{c|}{90.48} & 
  0.1501 &
  \multicolumn{1}{c|}{80.62} &
  0.005573 &
  \multicolumn{1}{c|}{77.93} & 
  0.4952 &
  \multicolumn{1}{c}{85.27} \\
  \multicolumn{1}{c|}{MaskNet \cite{masknet}} &
  \underline{0.3711} &
  \multicolumn{1}{c|}{\underline{79.43}} &
  0.4387 &
  \multicolumn{1}{c|}{81.34} &
  0.3080 &
  \multicolumn{1}{c|}{90.34} &
  0.1498 &
  \multicolumn{1}{c|}{80.79} &
  0.005556 &
  \multicolumn{1}{c|}{77.85} &
  0.5003 &
  \multicolumn{1}{c}{84.79}\\
  \multicolumn{1}{c|}{EulerNet \cite{EulerNet}} &
  0.3723 &
  \multicolumn{1}{c|}{79.22} &
  0.4379 &
  \multicolumn{1}{c|}{81.47} &
  0.3050 &
  \multicolumn{1}{c|}{90.44} &
  0.1498 &
  \multicolumn{1}{c|}{80.78} &
  0.005540 &
  \multicolumn{1}{c|}{\underline{78.30}} &
  0.4922 &
  \multicolumn{1}{c}{84.27}\\ 
  \multicolumn{1}{c|}{FinalMLP \cite{finalmlp}} &
  0.3718 &
  \multicolumn{1}{c|}{79.35} &
  0.4373 &
  \multicolumn{1}{c|}{81.45} &
  0.3058 &
  \multicolumn{1}{c|}{90.52} &
  \underline{0.1497} &
  \multicolumn{1}{c|}{80.78} &
  0.005556 &
  \multicolumn{1}{c|}{78.02} &
  0.4822 &
  \multicolumn{1}{c}{85.10}\\ 
  \multicolumn{1}{c|}{FINAL \cite{FINAL}} &
  0.3712 &
  \multicolumn{1}{c|}{79.41} &
  \underline{0.4371} &
  \multicolumn{1}{c|}{\underline{81.49}} &
  \underline{0.3035} &
  \multicolumn{1}{c|}{90.53} &
  0.1498 &
  \multicolumn{1}{c|}{80.74} &
  0.005540 &
  \multicolumn{1}{c|}{78.13} &
  0.4800 &
  \multicolumn{1}{c}{85.14} \\ 
  \multicolumn{1}{c|}{RFM \cite{RFM}} &
  0.3723 &
  \multicolumn{1}{c|}{79.24} &
  0.4374 &
  \multicolumn{1}{c|}{81.47} &
  0.3048 &
  \multicolumn{1}{c|}{90.51} &
  0.1506 &
  \multicolumn{1}{c|}{80.73} &
  0.005540 &
  \multicolumn{1}{c|}{78.25} &
  0.4853 &
  \multicolumn{1}{c}{84.70}\\
  \multicolumn{1}{c|}{DLF \cite{DLF}} &
  0.3720 &
  \multicolumn{1}{c|}{79.31} &
  0.4382 &
  \multicolumn{1}{c|}{81.40} &
  0.3083 &
  \multicolumn{1}{c|}{90.52} &
  \underline{0.1497} &
  \multicolumn{1}{c|}{\underline{80.81}} &
  0.005540 &
  \multicolumn{1}{c|}{78.09} &
  0.4884 &
  \multicolumn{1}{c}{85.07}\\ \hline
  \multicolumn{1}{c|}{\textbf{ECN}} &
  0.3698$^\star$ &
  \multicolumn{1}{c|}{79.68$^\star$} &
  0.4365$^\star$ &
  \multicolumn{1}{c|}{81.56$^\star$} &
  0.3023$^\star$ &
  \multicolumn{1}{c|}{90.67$^\star$} &
  0.1496 &
  \multicolumn{1}{c|}{80.88$^\star$} &
  0.005532$^\star$ &
  \multicolumn{1}{c|}{78.50$^\star$} &
  0.4756$^\star$ &
  \multicolumn{1}{c}{85.59$^\star$}\\
  \multicolumn{1}{c|}{\textbf{FCN$_p$}} &
  0.3697$^\star$ &
  \multicolumn{1}{c|}{79.66$^\star$} &
  0.4361$^\star$ &
  \multicolumn{1}{c|}{81.60$^\star$} &
  \textbf{0.2975$^\star$} &
  \multicolumn{1}{c|}{\textbf{90.82$^\star$}} &
  \textbf{0.1494$^\star$} &
  \multicolumn{1}{c|}{\textbf{80.99$^\star$}} &
  0.005534$^\star$ &
  \multicolumn{1}{c|}{78.48$^\star$} &
  \textbf{0.4747$^\star$} &
  \multicolumn{1}{c}{85.67$^\star$} \\
  \multicolumn{1}{c|}{\textbf{FCN$_{sp}$}} &
  \textbf{0.3693$^\star$} &
  \multicolumn{1}{c|}{\textbf{79.72$^\star$}} &
  \textbf{0.4357$^\star$} &
  \multicolumn{1}{c|}{\textbf{81.63$^\star$}} &
   0.3017$^\star$ &
  \multicolumn{1}{c|}{90.67$^\star$} &
   \textbf{0.1494$^\star$} &
  \multicolumn{1}{c|}{80.96$^\star$} &
  \textbf{0.005529$^\star$} &
  \multicolumn{1}{c|}{\textbf{78.52$^\star$}} &
  0.4754$^\star$ &
  \multicolumn{1}{c}{\textbf{85.74$^\star$}} \\ \hline
   \multicolumn{1}{c|}{\textit{Abs.Imp}} &
  -0.0018 &
  \multicolumn{1}{c|}{+0.29} &
  -0.0014 &
  \multicolumn{1}{c|}{+0.14} &
   -0.0060 &
  \multicolumn{1}{c|}{+0.26} &
   -0.0003 &
  \multicolumn{1}{c|}{+0.18} &
  -0.000010 &
  \multicolumn{1}{c|}{+0.22} &
  -0.0019 &
  \multicolumn{1}{c}{+0.39}\\
  \Xhline{1px}
\end{tabular}}
\label{implicit}
\end{table*}

\begin{figure*}[t]
   
  \centering
  \includegraphics[width=1\textwidth]{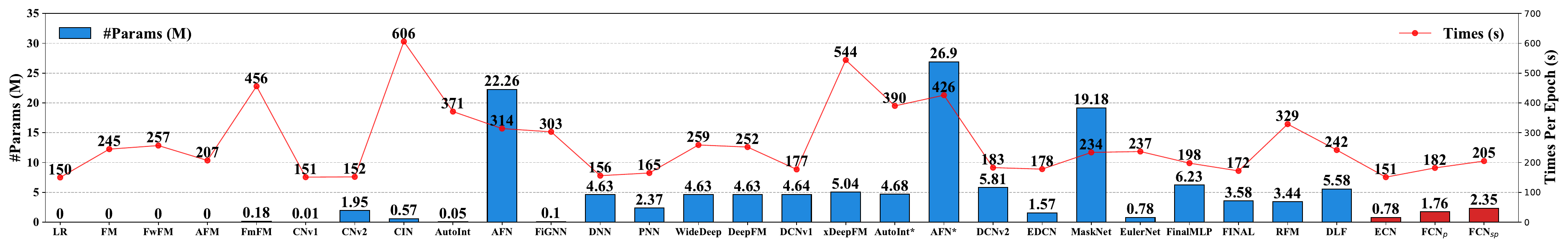}
   
  \captionsetup{justification=raggedright}
  \caption{\textbf{Efficiency comparisons with other models on the Criteo dataset. We only consider non-embedding parameters. We fix the optimal performance hyperparameters for each model and conduct experiments uniformly on one GeForce RTX 4090 GPU.}}
  \label{EfficiencyComparisons}
   
\end{figure*}

\subsubsection{\textbf{Evaluation Metrics.}} To compare the performance, we utilize two commonly used metrics in CTR models: \textbf{Logloss}, \textbf{AUC} \cite{autoint, GDCN, ComboFashion}. AUC stands for Area Under the ROC Curve, which measures the probability that a positive instance will be ranked higher than a randomly chosen negative one. A lower Logloss suggests a better capacity for fitting the data.

\subsubsection{\textbf{Baselines.}} To verify the superiority of ECN and FCN over models that include implicit feature interactions, we further select several representative baselines, such as PNN \cite{pnn1} and Wide \& Deep \cite{widedeep} (2016); DeepFM \cite{deepfm} and DCNv1 \cite{dcn} (2017); xDeepFM (2018) \cite{xdeepfm}; AutoInt* (2019) \cite{autoint}; AFN* (2020) \cite{AFN}; DCNv2 \cite{dcnv2} and EDCN \cite{EDCN}, MaskNet \cite{masknet} (2021); EulerNet \cite{EulerNet}, FinalMLP \cite{finalmlp}, FINAL \cite{FINAL} (2023), RFM \cite{RFM} (2024); DLF \cite{DLF} (2025).

\subsubsection{\textbf{Implementation Details.}} We implement all models using PyTorch \cite{PYTORCH} and refer to existing works \cite{openbenchmark, FuxiCTR}. We employ the Adam optimizer \cite{adam} to optimize all models, with a default learning rate set to 0.001. For the sake of fair comparison, we set the embedding dimension to 128 for KKBox and 16 for the other datasets \cite{openbenchmark, Bars}. The Dropout rate is determined via grid search over the set \{0, 0.1, 0.2, 0.3\}. The batch size is set to 4,096 on the ML-1M and iPinYou datasets and 10,000 on the other datasets. During training, we employ a Reduce-LR-on-Plateau scheduler that reduces the learning rate by a factor of 10 when performance stops improving in any given epoch \cite{openbenchmark,Bars}. To prevent overfitting, we employ early stopping with a patience value of 2. The hyperparameters of the baseline model are configured and fine-tuned based on the \textit{optimal values} provided in \cite{FuxiCTR,openbenchmark,Bars} and their original paper. For datasets not included in open-source baseline libraries \cite{FuxiCTR,openbenchmark,Bars}, we use a DNN architecture with [400,400,400] and apply the same hyperparameter search strategy as described in the original model papers. For models not covered in \cite{FuxiCTR}, we use the source code released by the authors. Further details on model hyperparameters and dataset configurations can be found in our running logs\footnote{\url{https://github.com/salmon1802/FCN/tree/KDD'26/checkpoints}}.

\subsection{Overall Performance}
 To further comprehensively investigate the performance superiority and generalization ability of FCN on various CTR datasets (e.g., large-scale sparse datasets), we select 16 representative baseline models and 6 benchmark datasets. We highlight the performance of ECN and FCN in bold and underline the best baseline performance. Table \ref{implicit} presents the experimental results, from which we can make the following observations:
\begin{itemize}[leftmargin=*]
\item Overall, FCN achieves the best performance across all six datasets, with an average AUC improvement of 0.25\% over the strongest baseline model and an average Logloss decrease of 0.19\%, both exceeding the statistically significant threshold of \textit{0.1\%}. This demonstrates the effectiveness of FCN. Besides, FCN$_p$ and FCN$_{sp}$ exhibit varying performance across different datasets. Therefore, we recommend flexibly adjusting the network architecture based on data distribution.
\item The FinalMLP model achieves good performance on the Avazu and Criteo datasets, surpassing most CTR models that combine explicit and implicit feature interactions. This demonstrates the effectiveness of implicit feature interactions. Consequently, most CTR models attempt to integrate DNN into explicit feature interaction models to enhance performance. However, FCN achieves SOTA performance using only explicit feature interactions, indicating the effectiveness and potential of modeling with explicit feature interactions alone.
\end{itemize}

\subsection{In-Depth Study of FCN}
\subsubsection{\textbf{Efficiency Comparison}}

To verify the efficiency of FCN, we fix the optimal hyperparameters of the 25 baseline models and compare their parameter count (rounded to two decimal places) and runtime (averaged over five runs). The experimental results are shown in Figure \ref{EfficiencyComparisons}. We can derive:
\begin{itemize}[leftmargin=*]
\item Explicit CTR models typically use fewer parameters. For instance, LR, FM, FwFM, and AFM have nearly zero non-embedding parameters, while FmFM, CrossNet, CIN, and AutoInt all require fewer than 1M parameters. Notably, parameter count does not always correlate with time complexity. Although CIN uses only 0.57M parameters, its training time per epoch reaches a maximum of 606 seconds, making it unsuitable for practical production environments. FiGNN and AutoInt face the same issue.
\item Compared with models deployed in production environments, such as FinalMLP, FINAL, and DCNv2, our proposed FCN requires fewer parameters while maintaining comparable training cost. With the introduction of the Tri-BCE loss, the training cost of FCN$_{sp}$ increases by only 31 seconds compared to ECN. Besides, the additional computational cost for the loss is incurred solely during training and does not impact inference speed. These results further demonstrate the efficiency of ECN and FCN.
\end{itemize}

\begin{table}[t]
\renewcommand\arraystretch{1}
\centering
\caption{\textbf{Ablation study of FCN$_{sp}$.}} 
\label{Ablation Study}
 
\resizebox{\linewidth}{!}{
\begin{tabular}{c|cc|cc|cc}
\Xhline{1px}
\multirow{2}{*}{\textbf{Model}}  & \multicolumn{2}{c|}{\textbf{Criteo}}  & \multicolumn{2}{c|}{\textbf{iPinYou}} & \multicolumn{2}{c}{\textbf{KKBox}} \\ \cline{2-7} 
                       & $ \text{Logloss}\downarrow$       & $ \text{AUC(\%)}\uparrow$        & 
                       $\text{Logloss}\downarrow$       & $ \text{AUC(\%)}\uparrow$ & $ \text{Logloss}\downarrow$       & $ \text{AUC(\%)}\uparrow$ \\ \hline
w/o LCN              & 0.4362   & 81.58 & 0.005534 & 78.48 & 0.4758 & 85.70  \\ 
w/o ECN              & 0.4367   & 81.53 & 0.005537 & 78.17 & 0.4826 & 85.39 \\ 
w/o TB         & 0.4367   & 81.55  & 0.005530 & 78.45 & 0.4777 & 85.56  \\ \hline 
FCN$_{sp}$                   & \textbf{0.4357}        & \textbf{81.63}  & \textbf{0.005529} & \textbf{78.52}  & \textbf{0.4754} & \textbf{85.74}    \\ \Xhline{1px}
\end{tabular}}
\label{ablation}
 
\end{table}

\begin{figure}[t]
    \subfloat[FCN$_p$ on Criteo]{
        \centering
        \includegraphics[width=0.5\linewidth]{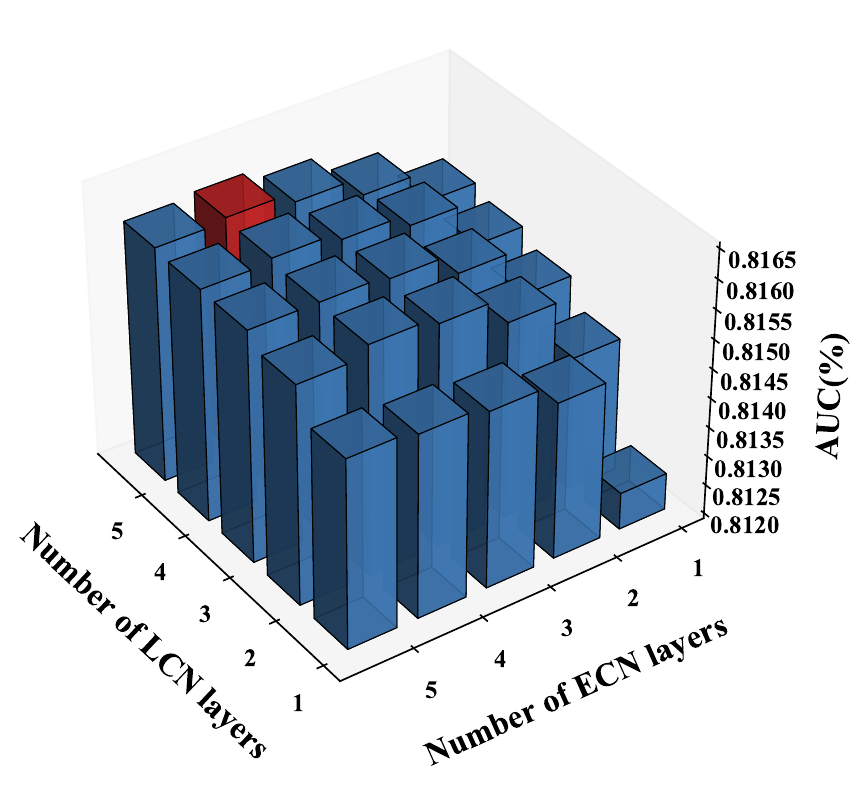}
    }
    \subfloat[FCN$_p$ on ML-1M]{
        \centering
        \includegraphics[width=0.5\linewidth]{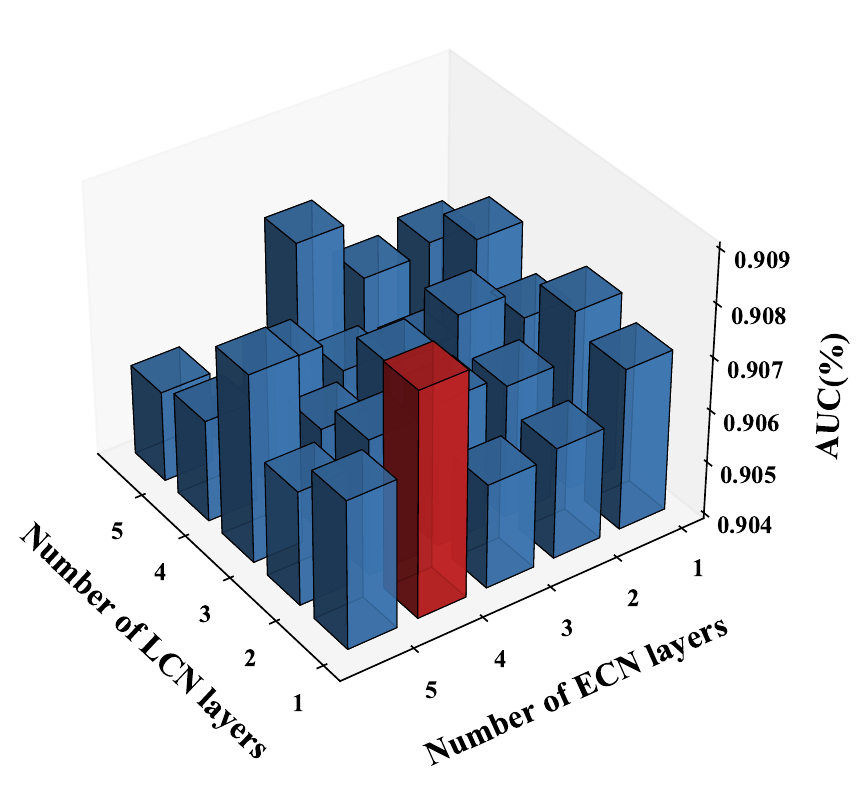}
    } \\  
    \subfloat[FCN$_{sp}$ on Criteo]{
        \centering
        \includegraphics[width=0.5\linewidth]{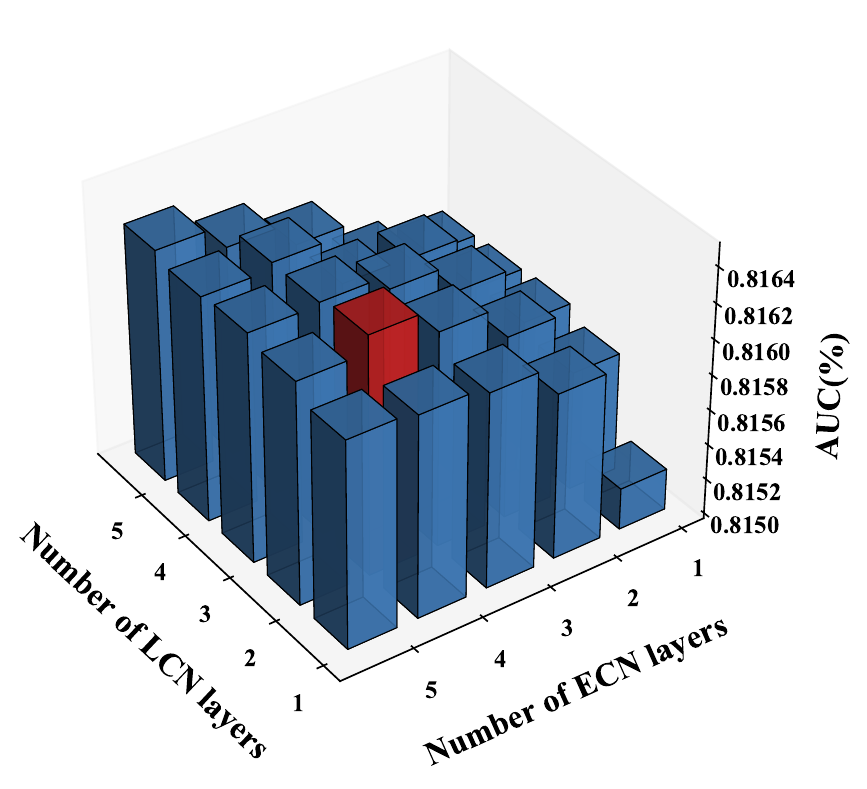}
    }
    \subfloat[FCN$_{sp}$ on ML-1M]{
        \centering
        \includegraphics[width=0.5\linewidth]{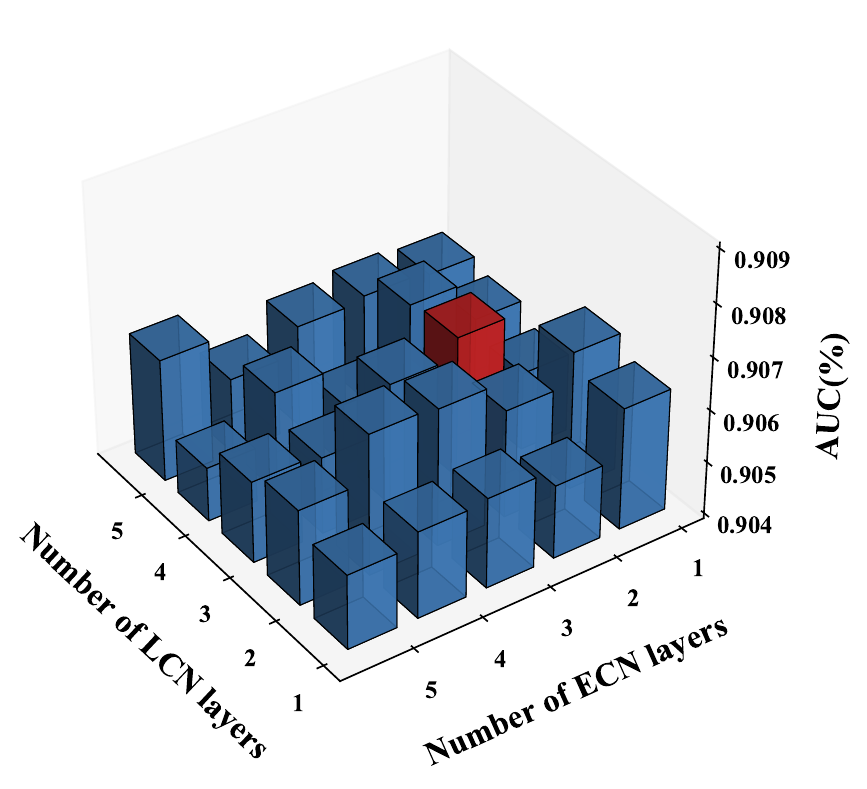}
    }  
    \captionsetup{justification=raggedright}
    \caption{Performance comparison for different network depths of FCN.}
    \label{layer_num}
 
\end{figure}

\subsubsection{\textbf{Ablation Study}}
To investigate the impact of each component of FCN$_{sp}$ on its performance, we conduct experiments on several variants using the three datasets where FCN$_{sp}$ achieves SOTA performance.
\begin{itemize}[leftmargin=*]
\item \textbf{w/o LCN}: FCN$_{sp}$ is constructed solely with ECN, while keeping the architecture and total number of layers unchanged.
\item \textbf{w/o ECN}: FCN$_{sp}$ is constructed solely with LCN, while keeping the architecture and total number of layers unchanged.
\item \textbf{w/o TB}: FCN$_{sp}$ with BCE instead of the Tri-BCE.
\end{itemize}
The results of the ablation experiments are presented in Table~\ref{ablation}. We observe that the \textbf{w/o LCN} variant results in the smallest performance degradation, while the \textbf{w/o ECN} variant leads to the largest performance drop. This indicates that ECN contributes more to the overall model performance than LCN. Meanwhile, the performance decrease of the \textbf{w/o LCN} variant also suggests that LCN and ECN serve as complementary interaction methods, employing a two-stream ECN yields suboptimal results. Besides, the variant w/o TB also leads to a certain degree of performance decline, particularly noticeable on KKBox. This further demonstrates the effectiveness of our proposed Tri-BCE.

\subsubsection{\textbf{Influence of Network Depths}} To further investigate the Influence of different neural network depths on the performance of FCN, we conduct experiments on Criteo and ML-1M datasets. From Figure \ref{layer_num}, we observe that the optimal layer configurations for FCN$_p$ and FCN$_{sp}$ differ. For instance, on the Criteo dataset, FCN$_p$ achieves optimal performance with a combination of 4 ECN layers and 5 LCN layer, whereas FCN$_{sp}$ performs best with 4 ECN layers and 2 LCN layers.

\subsubsection{\textbf{Influence of Low-cost Aggregation}}
To investigate the impact of LCA on model performance, we conduct experiments on the Criteo and KKBox datasets. The results are shown in Table~\ref{Influence_LCA}, where "-full" variant denotes a full-rank network without LCA. We observe that the "-full" variant does not demonstrate significant performance advantages and even suffers from performance degradation in some cases. Meanwhile, the "-full" variant requires twice as many network parameters as LCA and increases inference latency by approximately 23\%. These results indicate that LCA effectively reduces model complexity without sacrificing performance.

\begin{table}[t]
\renewcommand\arraystretch{1}
\centering
\caption{Influence of Low-cost Aggregation.}

\resizebox{\linewidth}{!}{
\begin{tabular}{c|cccc|cccc}
\Xhline{1px}
& \multicolumn{4}{c|}{\textbf{Criteo}} & \multicolumn{4}{c}{\textbf{KKBox}} \\ \cline{2-9} 
\multirow{-2}{*}{\textbf{Model}} & Logloss$\downarrow$  & AUC(\%)$\uparrow$    & Params$\downarrow$   & Latency$\downarrow$ & Logloss$\downarrow$  & AUC(\%)$\uparrow$   & Params$\downarrow$  & Latency$\downarrow$ \\ \hline
ECN & 0.4365 & 81.56 & \textbf{0.78M} & \textbf{2.85ms} & \textbf{0.4756} & 85.59 & \textbf{5.55M} & \textbf{13.54ms} \\
ECN-full & \textbf{0.4363}   & \textbf{81.57}  & 1.56M & 3.62ms & 0.4759 & \textbf{85.60} & 11.08M & 17.77ms   \\ \hline
FCN$_p$ & \textbf{0.4361} & \textbf{81.60} & \textbf{1.76M} & \textbf{5.11ms} & \textbf{0.4747} & \textbf{85.67} & \textbf{11.10M} & \textbf{27.15ms} \\
FCN$_p$-full & 0.4362   & 81.59 & 3.51M & 6.54ms  & 0.4786 & 85.63 & 22.16M & 34.41ms   \\ \Xhline{1px}
\end{tabular}}
\label{Influence_LCA}
 
\end{table}

\subsubsection{\textbf{The Performance Gap between ECN and LCN}}

\begin{figure}[t]
    \subfloat[ECN vs LCN on Criteo]{
        \centering
        \includegraphics[width=0.5\linewidth]{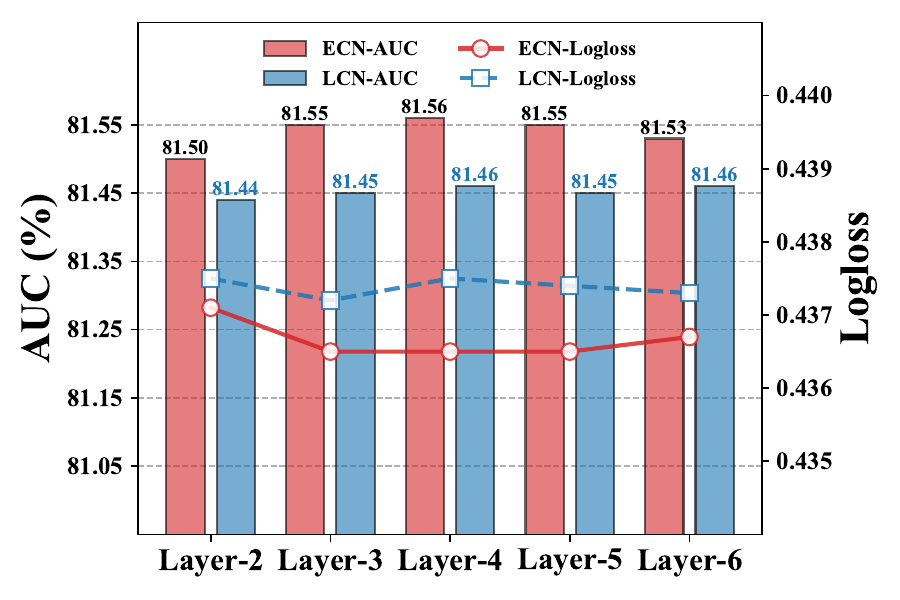}
    }
    \subfloat[ECN vs LCN on KKBox]{
        \centering
        \includegraphics[width=0.5\linewidth]{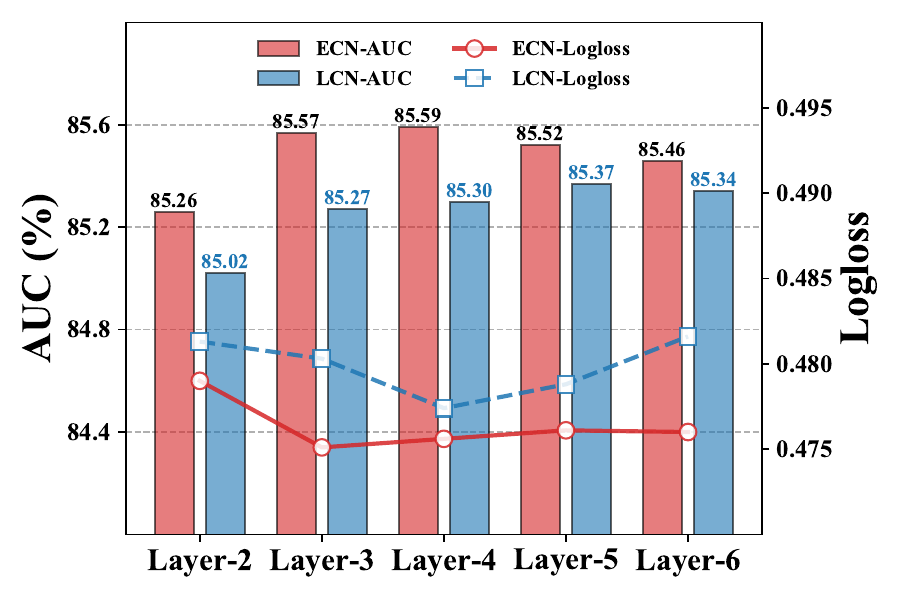}
    }
     
    \captionsetup{justification=raggedright}
    \caption{Performance comparison of ECN and LCN.}
    \label{gap_ecn_lcn}
 
\end{figure}
To investigate the performance gap between ECN and LCN, we conduct the experiments shown in Fig.~\ref{gap_ecn_lcn}. We observe that as the number of layers increases, ECN consistently outperforms LCN in terms of both AUC and Logloss. This further demonstrates the effectiveness of exponentially growing feature interaction methods.

\subsubsection{\textbf{Industrial Evaluation}}
To investigate the effectiveness of FCN in industrial-scale recommender systems, we conduct offline evaluations using real production user click logs collected over eight consecutive days (the first seven days serve as the training set, and the last day serves as the validation set). The experimental results are reported in Table \ref{Industrial}, which present the prediction performance of seven-day post-click conversion rate (CVR) \cite{ESMM} for clicked samples in two core business domains. Specifically, ECN+ is obtained from DCNv2 by modifying only a single variable in the code, namely replacing the anchor feature from $\boldsymbol{x}_{1}$ to $\boldsymbol{x}_{l}$. As shown in Table \ref{Industrial}, ECN+ performs slightly worse than DCNv2 in Domain 1, while achieving a notable improvement in CVR prediction performance in Domain 2. This indicates that Domain 1 may not require high-order feature interactions, in which case CrossNetv2 suffices, whereas Domain 2 requires higher-order feature interactions to improve model performance.

\begin{table}[t]
\renewcommand\arraystretch{1}
\centering
\caption{Offline results in production settings.}
\resizebox{\linewidth}{!}{
\begin{tabular}{cc|ccccccc}
\Xhline{1px}
\multicolumn{2}{c|}{AUC}          & Day1   & Day2   & Day3   & Day4   & Day5   & Day6   & Day7   \\ \hline
\multirow{2}{*}{Domain 1} & DCNv2 & \textbf{0.8609} & \textbf{0.9071} & \textbf{0.9209} & \textbf{0.9252} & \textbf{0.9279} & \textbf{0.9250} & \textbf{0.9227} \\
                          & ECN*   & 0.8553 & 0.9054 & 0.9193 & 0.9229 & 0.9272 & 0.9228 & 0.9170 \\ \hline
\multirow{2}{*}{Domain 2} & DCNv2 & 0.8414 & 0.8181 & 0.8745 & 0.8454 & 0.8119 & 0.8511 & 0.8391 \\
                          & ECN*   & \textbf{0.8647} & \textbf{0.8493} & \textbf{0.8880} & \textbf{0.8921} & \textbf{0.8961} & \textbf{0.8815} & \textbf{0.8773} \\ \Xhline{1px}
\end{tabular}}
\vspace{-1em}
\label{Industrial}
\end{table}

\section{Conclusion}
This paper introduced the next generation deep cross network, called FCN, which uses sub-networks LCN and ECN to capture both low-order and high-order feature interactions without relying on the less interpretable DNN. LCN uses a linearly growing interaction method for low-order interactions, while ECN employs an exponentially increasing method for high-order interactions. The low-cost aggregation further improves FCN’s computational efficiency. Tri-BCE helped the two sub-networks in FCN obtain more suitable supervision signals for themselves. Comprehensive experiments on six datasets demonstrated the effectiveness and efficiency of FCN.

\begin{acks}
This work is supported by the National Science Foundation of China (No. 62272001 and No. 62206002).
\end{acks}

\bibliographystyle{ACM-Reference-Format}
\bibliography{sample-base}










\end{document}